\documentclass[11pt,letterpaper]{article}

\usepackage{jheppub_mod}
\usepackage{slashed}
\usepackage{graphicx}
\usepackage{psfrag}
\usepackage{pstricks}
\usepackage{calc}

\makeatletter
\newsavebox\myboxA
\newsavebox\myboxB
\newlength\mylenA

\newcommand*\xoverline[2][0.75]{%
    \sbox{\myboxA}{$\m@th#2$}%
    \setbox\myboxB\null% Phantom box
    \ht\myboxB=\ht\myboxA%
    \dp\myboxB=\dp\myboxA%
    \wd\myboxB=#1\wd\myboxA% Scale phantom
    \sbox\myboxB{$\m@th\overline{\copy\myboxB}$}%  Overlined phantom
    \setlength\mylenA{\the\wd\myboxA}%   calc width diff
    \addtolength\mylenA{-\the\wd\myboxB}%
    \ifdim\wd\myboxB<\wd\myboxA%
       \rlap{\hskip 0.5\mylenA\usebox\myboxB}{\usebox\myboxA}%
    \else
        \hskip -0.5\mylenA\rlap{\usebox\myboxA}{\hskip 0.5\mylenA\usebox\myboxB}%
    \fi}
\makeatother

\RequirePackage{color}

\newcommand{\cA}{\mathcal{A}}

\newcommand{\cM}{\mathcal{M}}
\newcommand{\cN}{\mathcal{N}}
\newcommand{\cO}{\mathcal{O}}

\newcommand{\cS}{\mathcal{S}}
\newcommand{\cU}{\mathcal{U}}
\newcommand{\cV}{\mathcal{V}}
\newcommand{\cW}{\mathcal{W}}

\newcommand{\cY}{\mathcal{Y}}

\newcommand{\uD}{\mathrm{D}}
\newcommand{\uF}{\mathrm{F}}

\newcommand{\ud}{\mathrm{d}}
\newcommand{\ue}{\mathrm{e}}
\newcommand{\ui}{\mathrm{i}}

\newcommand{\us}{\mathrm{s}}

\newcommand{\II}{\mathbb{I}}
\newcommand{\RR}{\mathbb{R}}
\newcommand{\ZZ}{\mathbb{Z}}

\newcommand{\Ga}{\Gamma}

\newcommand{\ep}{\epsilon}
\newcommand{\ga}{\gamma}
\newcommand{\ka}{\kappa}

\newcommand{\sig}{\sigma}

\newcommand{\vp}{\varphi}

\newcommand{\vol}{\mathrm{vol}}

\newcommand{\SO}[1]{\mathrm{SO}\left(#1\right)}
\newcommand{\SU}[1]{\mathrm{SU}\left(#1\right)}
\newcommand{\U}[1]{\mathrm{U}\left(#1\right)}

\newcommand{\tr}{\mathrm{tr}\,}

\title{On Chiral Mesons in AdS/CFT}

\author{Liam McAllister, Paul McGuirk, and John Stout}

\affiliation{Department of Physics,
Cornell
  University, Ithaca, NY 14853, USA}

\emailAdd{mcallister@cornell.edu}
\emailAdd{mcguirk@cornell.edu}
\emailAdd{jes554@cornell.edu}

\abstract{We analyze the spectra of non-chiral and chiral
  bifundamental mesons arising on intersecting D7-branes in
  $AdS_{5}\times S^{5}$.  In the absence of magnetic flux on the curve
  of intersection, the spectrum is non-chiral, and the dual gauge theory
  is conformal in the quenched/probe approximation.  For this case we
  calculate the dimensions of the bifundamental mesonic operators. We
  then consider magnetization of the D7-branes, which deforms the dual
  theory by an irrelevant operator and renders the mesons chiral.
  The magnetic flux spoils the conformality of the dual
  theory, and induces a D3-brane charge that becomes large in the
  ultraviolet, where the non-normalizable bifundamental modes are rapidly
  divergent.  An ultraviolet completion is therefore necessary
  to calculate the correlation functions in the chiral case.
  On the other hand, the normalizable modes are very well
  localized in the infrared, leading to new possibilities for local
  model-building on intersecting D7-branes in warped geometries.}

\begin{document}

\maketitle

%%%%
%%%%
\section{\label{sec:intro}Introduction}
%%%%
%%%%

The AdS/CFT
correspondence~\cite{Maldacena:1997re,Witten:1998qj,Gubser:1998bc,Aharony:1999ti}
is a powerful duality relating conformal field theories (CFTs) in $d$
dimensions to gravitational theories on $\left(d+1\right)$-dimensional
anti-de Sitter (AdS) spaces.  The extension of the duality to include
global flavor groups has been well-studied
(see~\cite{Aharony:1998xz,Karch:2002sh} for some foundational work)
and is well-motivated: it brings the theory closer to
phenomenologically viable models, with mesonic bound states serving as
prototypes for visible-sector fields.  However, to find more realistic
models, the flavor group must be extended to a product group, and the
resulting mesonic spectrum must be made chiral.  Such extensions have
been relatively unexplored, and in the present work we report on
progress in this direction.

When the gravity side of the duality is a type II string theory,
flavor groups are added through the introduction of higher-dimensional
D$p$-branes that fill AdS and wrap compact
cycles~\cite{Karch:2002sh}.\footnote{The higher-dimensional D-brane
  need not fill all of AdS; if the brane is characterized by a minimum
  distance away from the origin of AdS then the dual quarks are
  massive~\cite{Kruczenski:2003be}.}  The simplest such example is the
addition of $F$ D7-branes to type IIB string theory on $AdS_{5}\times
S^{5}$, where we take the D7-branes to fill $AdS_{5}$ and wrap an
$S^{3}$ of the $S^{5}$.  The geometry is supported by $N$ units of
D3-brane charge and, without the D7-branes, is dual to $\cN=4$
$\SU{N}$ super Yang-Mills.  Adding the D7-branes deforms the dual
theory to an $\cN=2$ gauge theory with a $\U{F}$ flavor group,
containing a massless adjoint hypermultiplet as well as a massless
quark hypermultiplet that transforms in the bifundamental of $\SU{N}
\times \U{F}$.  The brane construction makes this clear, as the open
string excitations of the D7-branes give rise to a $\U{F}$ gauge
theory, and the infinite D7-brane worldvolume transverse to the
D3-branes results in a vanishing 4d coupling for this theory.  Open
strings stretching between the D7-branes and the D3-branes have the
same charges as the quarks in the dual theory.  We will work in the
standard decoupling limit~\cite{Maldacena:1997re} in
which one first takes
\begin{equation}
  g_{\us} \to 0, \qquad
  N \to \infty, \qquad
  \lambda_t \equiv 4\pi g_{\us} N \quad {\rm{fixed}},
  \label{decouplinglimit}
\end{equation}
and then sends the 't~Hooft coupling $\lambda_t$ to infinity.  In this
limit, the D3-branes are replaced by their near-horizon backreaction,
so that the only open strings are those stretching among the
D7-branes.  These transform in the adjoint representation of $\U{F}$
and are dual to mesonic operators in the gauge theory.

D7-branes are codimension-two objects, and so their backreaction cannot
generally be neglected.  Correspondingly, the presence of quarks in
the dual gauge theory alters the renormalization group flow, which was trivial before the
introduction of flavor.  Fortunately, the decoupling limit
(\ref{decouplinglimit})
simplifies the situation: if we
hold fixed the number of flavors, $F$, while taking the number of
colors to be large, then one can consistently neglect the running of
quarks in loops.  In the dual geometry, many aspects of the D7-brane
backreaction scale as $F/N$ and so also vanish in this limit
(see~\cite{Nunez:2010sf} and references therein). The flavored gauge
theory does have a Landau pole, and so the influence of the quarks on
the renormalization group flow cannot be neglected forever, but the scale at which the Landau
pole appears grows exponentially with $N/F$.  This so-called quenched
approximation, in which the running of quarks in
loops
is
neglected, is equivalent to the limit in which the D7-branes are taken
as
probes
of the dual geometry.  In what follows, we will take
this approximation without further apology.

The introduction of flavor branes opens up significant possibilities
for model-building.  Dimensional reduction along the angular
directions provides a framework for Randall-Sundrum
constructions~\cite{Randall:1999ee,Verlinde:1999fy,Greene:2000gh,Giddings:2001yu}
wherein the Standard Model fields propagating in the
bulk~\cite{Davoudiasl:1999tf,*Pomarol:1999ad,*Chang:1999nh,*Grossman:1999ra,*Gherghetta:2000qt}
descend from the D7-brane fluctuations as in~\cite{Gherghetta:2006yq}.
Upon compactification, the flavor group on the D7-branes becomes a
prototype for the Standard Model gauge group.
Of course, the Standard Model gauge group is a product; a corresponding
product flavor group results from introducing two
separate stacks of D7-branes.  The bifundamental fields are then
open strings stretching between the stacks, and in order for some of the
bifundamentals to be massless, the stacks must intersect.

A further challenge is that the Standard Model spectrum is chiral.  In
the class of constructions considered here, chirality in the 4d theory
can be induced by introducing magnetic flux on the (noncompact) curve
where the D7-branes intersect.  Upon compactification to 4d, the zero
modes of the Dirac operator acquire a net chirality set by the amount
of quantized magnetic flux.

Yet another difficulty in embedding fully realistic theories into
warped backgrounds of string theory is the fact that the Standard
Model is not a supersymmetric theory.  In geometries that are
characterized by a finite infrared scale, such as the well-studied
Klebanov-Strassler solution~\cite{Klebanov:2000hb}, supersymmetry can
be broken in a controllable way by the addition of a small number of
anti-D3-branes~\cite{Kachru:2003aw}.  The resulting
geometry~\cite{Bena:2009xk,DeWolfe:2008zy,*McGuirk:2009xx,*Dymarsky:2011pm,*Bena:2011hz,*Bena:2011wh,*Massai:2012jn,*Bena:2012bk,*Dymarsky:2013tna}
corresponds to the spontaneous breaking of supersymmetry in the dual
field theory~\cite{Kachru:2002gs}.\footnote{Some authors have
  interpreted the singularities of the anti-D3-brane geometry described
  in~\cite{Bena:2009xk} as implying that the supersymmetry-breaking
  state does not exist.}  An alternative is to consider ``gluing'' the
warped geometry to a compact space that does not preserve
supersymmetry.  The dual field theory is then a non-supersymmetric
theory with emergent supersymmetry, as
in~\cite{Gherghetta:2003he,*Sundrum:2009gv}.  Although
non-supersymmetric constructions are difficult to control,
the filtering provided by the renormalization group means that the
influence of the non-supersymmetric bulk, including the effects of
moduli stabilization, can be systematically parameterized and
incorporated along the lines of~\cite{Baumann:2010sx,*Gandhi:2011id}.
No matter which supersymmetry-breaking mechanism is
used,\footnote{See~\cite{Kachru:2009kg,*Dymarsky:2011ve} for other
  interesting proposals.} the resulting geometry is considerably more
complex after supersymmetry is broken.  We will therefore, in this
initial work, focus on supersymmetric D7-brane probes of
supersymmetric backgrounds.\footnote{See, for
  example,~\cite{Camara:2004jj,*Lust:2004fi,*Lust:2004dn,*Burgess:2006mn,*Lust:2008zd,*McGuirk:2009am,*McGuirk:2011yg,Benini:2009ff,McGuirk:2012tv}
  for analyses of probe D7-branes in non-supersymmetric backgrounds
  from the worldvolume and/or worldsheet points of view.}

In this note, we will consider the non-chiral and chiral bifundamental
modes existing at the intersections of probe D7-branes in
$AdS_{5}\times S^{5}$.  We build up to the chiral, warped case through
the simpler example of intersecting D7-branes in flat space
(\S\ref{sec:unwarped}).  Although the flat-space analysis of
\S\ref{sec:unwarped} has appeared elsewhere in the literature (see e.g.~\cite{Hashimoto:2003xz,*Nagaoka:2003zn,Cecotti:2009zf,Marchesano:2010bs}),
a detailed treatment is useful here, because the equations of motion
are readily generalized from the simple flat-space case to the
$AdS_{5}\times S^{5}$ configuration of primary interest.

The organization of this note is as follows.  In \S\ref{sec:unwarped}
we begin with the simple case of intersecting D7-branes in a flat
space background.  In \S\ref{flatspacenonchiral} we compute the mass
spectrum of the bifundamental modes for the case of vanishing magnetic
flux, where the spectrum is non-chiral.  Then, in
\S\ref{flatspacechiral} we calculate the chiral mass spectrum in a
configuration with magnetic flux.  Next, in \S\ref{sec:vec_mesons} we
consider unmagnetized intersecting D7-branes in $AdS_{5}\times S^{5}$,
computing the scaling dimensions of vector-like bifundamental mesonic
operators.  Finally, in \S\ref{sec:chiral_mesons} we add the simplest
possible magnetization to the intersecting D7-branes in $AdS_{5}\times
S^{5}$, and show that this magnetization makes the calculation of
correlation functions untrustworthy without an ultraviolet completion.
Concluding remarks are given in \S\ref{sec:conc}, while our
conventions and a few technical details appear in the appendices.

\newpage

%%%%
%%%%
\section{\label{sec:unwarped}D7-branes in Flat Space}
%%%%
%%%%

As a warm-up to the case of strong warping, we will first review the
case of intersecting D7-branes probing unwarped flat space,
$\mathbb{R}^{9,1} = \mathbb{R}^{3,1}\times \mathbb{C}^{3}$.

\subsection{The  D7-brane action}

As discussed in the introduction, we focus on supersymmetric
configurations, and so we take a flat D7-brane probe, which preserves
half of the supercharges of flat space.  By a choice of orientation
and complex structure, the D7-brane worldvolume $\cW$ can be taken to
be $\mathbb{R}^{3,1}\times \mathbb{C}^{2}$. The light bosonic degrees
of freedom resulting from the open-string excitations of the D7-brane
consist of the transverse deformations $\Phi^{i}$ and a $\U{1}$ vector
potential $A_{1}$.  We use this potential to construct a
Lorentz-invariant\footnote{Here and throughout we will use ``Lorentz
  invariance'' to refer to $\SO{3,1}$ invariance.}  and supersymmetric
magnetic flux $F_{2}=\ud A_{1}$; such a flux satisfies the
self-duality
condition~\cite{Marino:1999af,*Gomis:2005wc,*Martucci:2005ht}
\begin{equation}
  F_{2}=\tilde{\ast}_{4} F_{2},  \label{Hodgecondition}
\end{equation}
where $\tilde{\ast}_{4}$ is the Hodge star built from the metric on
$\mathbb{C}^{2}$.  The condition (\ref{Hodgecondition}) is equivalent
to $F_{2}$ being $\left(1,1\right)$ and primitive with respect to the
K\"ahler form induced on $\mathbb{C}^{2}$.

To leading order in the $\alpha^{\prime}$ expansion, the action of the
D7-brane in this background
is~\cite{Marolf:2003ye,*Marolf:2003vf,*Martucci:2005rb}
\begin{align}
  S_{\uD 7}=&-\frac{1}{g_{8}^{2}}\int_{\cW}\!\!\ud^{8}\xi^{\alpha}\,\sqrt{-\hat{g}}
  \biggl\{\frac{1}{2}\hat{g}_{ij}\hat{g}^{\alpha\beta}
  \partial_{\alpha}\Phi^{i}\partial_{\beta}\Phi^{j}
  +\frac{1}{4}\hat{g}^{\alpha\beta}\hat{g}^{\ga\delta}
  F_{\alpha\ga}F_{\beta\delta}
  +\ui\bar{\Theta}P_{-}^{\uD 7}
  \hat{g}^{\alpha\beta}\hat{\Ga}_{\alpha}
  \partial_{\beta}\Theta\biggr\},
\end{align}
in which we have omitted a constant term that does not play a role in
our analysis.  Writing the string tension as $\tau_{\uF
  1}^{-1}=2\pi\alpha'=\ell_{\us}^{2}$, the 8d Yang-Mills coupling is
$g_{8}^{-2}=8\pi^{3}\ell_{\us}^{4}g_{\us}$.  Here $\xi^{\alpha}$ are
coordinates on the D7-brane, $\hat{g}_{\alpha\beta}$ is the induced
worldvolume metric and $\hat{g}_{ij}$ is the transverse metric.
$\Theta$ is a 10d double Majorana-Weyl spinor (reviewed in
Appendix~\ref{app:conv}) that, as in the Green-Schwarz superstring,
redundantly encapsulates the fermionic degrees of freedom of the
D7-brane.  In particular, $\Theta$ is subject to the $\ka$-symmetry
identification
\begin{equation}
  \Theta\sim\Theta+P_{-}^{\uD p}\ka,
\end{equation}
in which $\ka$ is an arbitrary Majorana-Weyl double spinor.
$P_{-}^{\uD 7}$ is given by
\begin{equation}
\label{eq:D7_proj}
  P_{-}^{\uD 7}=\frac{1}{2}
  \begin{pmatrix}
    1 & - \Ga^{-1}_{\uD 7} \\ -\Ga_{\uD 7} & 1
  \end{pmatrix},
\end{equation}
in which
\begin{equation}
  \Ga_{\uD 7}=\ud\slashed{\vol}_{\cW}:=
  \frac{1}{8!}\hat{\ep}_{\alpha_{1}\cdots\alpha_{8}}
  \hat{\Ga}^{\alpha_{1}\cdots\alpha_{8}}=-\ui\Ga_{\left(8\right)},
\end{equation}
where $\ep_{\alpha_{1}\cdots\alpha_{8}}$ is the antisymmetric tensor
and $\Ga_{\left(8\right)}$ is the $\SO{7,1}$ chirality operator.  We
use $\ka$-symmetry to set
\begin{equation}
  \label{eq:kappa-gauge}
  \Theta=\begin{pmatrix}\theta \\ 0\end{pmatrix}.
\end{equation}
With this choice,
\begin{equation}
  S_{\uD 7}=-\frac{1}{g_{8}^{2}}\int_{\cW}\ud^{8}\xi^{\alpha}\,
  \biggl\{\frac{1}{2}\hat{g}_{ij}\hat{g}^{\alpha\beta}
  \partial_{\alpha}\Phi^{i}\partial_{\beta}\Phi^{j}
  +\frac{1}{4}\hat{g}^{\alpha\beta}\hat{g}^{\ga\delta}
  F_{\alpha\ga}F_{\beta\delta}
  +\frac{\ui}{2}\bar{\theta}
  \hat{g}^{\alpha\beta}\hat{\Ga}_{\beta}
  \partial_{\alpha}\theta\biggr\},
\end{equation}
which is the familiar action for maximally supersymmetric 8d $\U{1}$
gauge theory.

On a stack of $F$ such D7-branes, the gauge group is enhanced to
$\U{F}$, and $A_{\alpha}$, $\Phi^{i}$, and $\theta$ are
promoted to adjoint-valued fields.
The leading-order action is determined by gauge-invariance
and supersymmetry to be
\begin{align}
  S_{\uD 7}=-\frac{1}{g_{8}^{2}}\int_{\cW}\ud^{8}\xi^{\alpha}\,
  \tr\biggl\{&\frac{1}{2}\hat{g}_{ij}\hat{g}^{\alpha\beta}
  D_{\alpha}\Phi^{i}D_{\beta}\Phi^{j}
  +\frac{1}{4}\hat{g}^{\alpha\beta}\hat{g}^{\ga\delta}
  F_{\alpha\ga}F_{\beta\delta}
  -\frac{1}{4}\hat{g}_{ij}\hat{g}_{kl}\bigl[\Phi^{i},\Phi^{k}\bigr]
  \bigl[\Phi^{j},\Phi^{l}\bigr]\notag\\
  &+\frac{\ui}{2}\bar{\theta}
  \hat{g}^{\alpha\beta}\hat{\Ga}_{\alpha}
  D_{\beta}\theta
  -\frac{1}{2}\bar{\theta}\,\hat{\Ga}_{i}
  \bigl[\Phi^{i},\theta\bigr]
  \biggr\},
  \label{eq:8d_sym}
\end{align}
in which $\mathrm{tr}$ denotes a trace over gauge indices, $D_{\alpha}$ is a
gauge covariant derivative
\begin{equation}
  D_{\alpha}=\partial_{\alpha}-\ui\bigl[A_{\alpha},\cdot\bigr],
\end{equation}
and $F_{2}=\ud A-\ui A\wedge A$ is the non-Abelian field strength.

Bifundamental modes arise from strings that stretch between stacks of
$\uD p$-branes.  If the stacks are parallel, then the mass of these
modes is proportional to the separation between the branes.  Such a
configuration still preserves sixteen supercharges and so the action
for the bifundamental modes (which provide a full massive vector
multiplet) can be fixed by symmetries.  Alternatively, the action can
be found by Higgsing the theory~\eqref{eq:8d_sym}.  Beginning with a
stack of $F_{1}+F_{2}$ D7-branes, the transverse deformations can be
treated as $\left(F_{1}+F_{2}\right)\times\left(F_{1}+F_{2}\right)$
matrices with the $i$th diagonal element corresponding to a transverse
deformation of the $i$th brane.  A vacuum expectation value (vev) with
the gauge structure
\begin{equation}
  \bigl\langle\Phi^{i}\bigr\rangle
  =\ell_{\us}^{-2}\begin{pmatrix}
    X_{1}^{i}\, \II_{F_{1}} & \\ &  X_{2}^{i}\, \II_{F_{2}}
  \end{pmatrix}
\end{equation}
breaks $\U{F_{1}+F_{2}}\to\U{F_{1}}\times \U{F_{2}}$ and describes a
separation of the branes $\Delta
x^{i}=\left\lvert X_{1}^{i}-X_{2}^{i}\right\rvert$. The factor of
$\ell_{\us}^{2}$ is introduced so that $\Phi^{i}$ has length dimension
$-1$.  However, this also has the effect of canceling the factors of
$\ell_{\us}$ that appear in operators correcting the Yang-Mills
action.  Therefore, in order to trust this effective field theory, we
consider cases where $\Delta x^{i}\ll \ell_{\us}$.  Equivalently, if
we are to trust the effective field theory description of the modes
stretching between the branes, their mass must be less than that of
the massive string states that have been integrated out implicitly.

Writing the fluctuations as
\begin{equation}
  \label{eq:fluc}
  \delta\Phi^{i}=\begin{pmatrix}
    \phi_{1}^{i} & \phi_{+}^{i} \\ \phi_{-}^{i} & \phi_{2}^{i}
  \end{pmatrix},
\end{equation}
$\phi_{1}^{i}$ and $\phi_{2}^{i}$ transform as adjoints under
$\U{F_{1}}$ and $\U{F_{2}}$, respectively, while $\phi_{+}^{i}$ and
$\phi_{-}^{i}$ are bifundamentals that acquire masses proportional to
the separation.  For notational simplicity, in what follows we will
consider the case $F_{1}=F_{2}=1$, but all of our results generalize
easily to higher ranks.

If, instead of being parallel, the branes intersect, some of the
bifundamental modes will become massless.  The intersection of two
D7-branes is generically six-dimensional, and the long-wavelength
description of the bifundamental modes can be given in terms of a 6d
effective field theory description on this intersection.  The 6d
masses of the bifundamentals depend on the angles formed by the
intersection of the branes.  However, the vector bifundamentals never
become massless, indicating that the 6d theory is a $\U{1}\times\U{1}$
(rather than the un-Higgsed $\U{2}$) gauge theory, and that fewer than
sixteen supercharges are preserved, since the vector multiplet is
split.  When the intersection is such that both D7-branes fill
$\mathbb{R}^{3,1}$ and are holomorphically embedded into
$\mathbb{C}^{3}$, at least minimal supersymmetry is
preserved~\cite{Berkooz:1996km} and the 6d theory includes massless
scalars and fermions.

\subsection{Non-chiral modes}  \label{flatspacenonchiral}

In the warped case, the calculation of mass spectra is equivalent to
the calculation of scaling dimensions in the dual theory.  In this
section, we continue our warm-up to the warped case by finding the
mass spectrum of non-chiral bifundamental modes in flat space.  To
this end, we take $z^{I=1,2,3}$ as coordinates on $\mathbb{C}^{3}$ and
consider a pair of D7-branes whose embeddings are specified by
\begin{equation}
  \uD 7_{1}: z^{3}=tz^{2},\quad
  \uD 7_{2}: z^{3}=-tz^{2},\qquad t>0.
\end{equation}
Following the discussion in the previous subsection, we can describe
this intersection by considering 8d $\U{2}$ SYM along
$\mathbb{R}^{3,1}\times \mathbb{C}^{2}$ (with $\mathbb{C}^{2}$ spanned
by $z^{1}$ and $z^{2}$), where the vev for the complexified transverse
deformation takes the form\footnote{Similar {vevs} were utilized
  in~\cite{Hashimoto:2003xz,*Nagaoka:2003zn} to describe brane
  recombination from non-supersymmetric intersections.}
\begin{equation}
  \label{eq:Higgs}
  \Phi=q\begin{pmatrix} z^{2} & \\ & -z^{2}\end{pmatrix},
\end{equation}
in which $q=\ell_{\us}^{-2}t$. The bifundamental modes are localized
on $\mathbb{R}^{3,1}\times \mathbb{C}$, with $z^{1}$ the coordinate on
the curve of intersection (which in this case is simply
$\mathbb{C}$). For the reasons discussed above, we must take $t\ll 1$
in order to trust the effective field theory.  Of course, no matter
what the value of $t$, at sufficiently large values of $z^{2}$ the
branes will be far apart and so one might worry about stringy
corrections to the Yang-Mills action.  That is, in addition
to~\eqref{eq:8d_sym}, the worldvolume action contains, for example,
operators with the schematic form
\begin{equation}
  \ell^{k-4}_{\us}\left(\Phi\right)^{k}\sim
  \left(\frac{t z^{2}}{\ell_{\us}}\right)^{k-4}\vp_{\pm}^{4}+\cdots,
\end{equation}
which
might
seem to become important at $z^{2}\sim t^{-1}\ell_{\us}$.
However, as we will show below, the bifundamental modes are highly
peaked at $z^{2}=0$, and so we anticipate that their physics will be
largely insensitive to the corrections at large $z^{2}$.

The configuration just described is supersymmetric, so we can find
solutions to the bosonic equations of motion by solving the fermionic
equations of motion.  Although the intersection is $\SO{5,1}$
symmetric, in anticipation of the magnetization --- which preserves only
$\SO{3,1}$ and which we discuss below --- we will make use of the
decomposition $\SO{9,1}\to\SO{3,1}\times\SO{6}$, as discussed in
Appendix~\ref{app:conv}.  It is useful to decompose the 10d fermionic
mode $\theta$ into modes of different internal chirality
(i.e. $\SO{6}$ weights)
\begin{equation}
  \label{eq:chirality_split}
  \theta=\sum_{m=0}^{3}\biggl\{\psi_{m}\begin{pmatrix}
    \xi \\ 0\end{pmatrix}
  \otimes\eta_{m}-
  \psi_{m}^{\dagger}\begin{pmatrix}
    0 \\ \sig^{2}\xi^{\ast}\end{pmatrix}\otimes
  \tilde{\beta}_{6}\eta_{m}^{\ast}\biggr\},
\end{equation}
in which $\xi$ is a fixed two-component spinor, $\eta_{m}$ are the
constant $\SO{6}$ positive chirality spinors in~\eqref{eq:pos_basis},
and $\tilde{\beta}_{6}$ is the $\SO{6}$ Majorana matrix.  Writing the
$\U{1}$ potential as $A_{1}=A_{\mu}\ud
x^{\mu}+\sum_{a=1}^{2}\bigl(a_{a}\ud
z^{a}+a_{\bar{a}}\ud\bar{z}^{a}\bigr)$, $\psi_{0}$ is the fermionic
partner of $A_{\mu}$, $\psi_{1,2}$ are the partners of $a_{1}$, and
$a_{2}$, and $\psi_{3}$ is the partner of the complexified transverse
deformation $\Phi$.  Each of the $\psi_{m}$ transforms under the
adjoint representation of $\U{2}$ and we write (cf.~\eqref{eq:fluc})
\begin{equation}
  \label{eq:bifund_decompose}
  \psi_{m}=
  \begin{pmatrix}
    & \psi_{m}^{+} \\ \psi_{m}^{-} &
  \end{pmatrix},
\end{equation}
in which we have set the neutral fields $\psi_{m}^{1,2}$ to zero since
they are not the modes of interest.

The linearized equation of motion for the fermions in this background
is
\begin{equation}
  0=\hat{\Ga}^{\alpha}\partial_{\alpha}\theta
  -\ui\,\hat{\Ga}_{i}\bigl[\Phi^{i},\theta\bigr],
\end{equation}
where the transverse fluctuations $\Phi^{i}$ are evaluated on their
vev~\eqref{eq:Higgs}. From $\SO{3,1}$ invariance, we expect that the
equation of motion for $A_{\mu}$ should decouple from those of the
other bosonic fields, at least for some gauge choice,\footnote{One
  such gauge choice is~\eqref{eq:D-term} after simply replacing the
  fermionic fields with their bosonic partners.  See, for
  example,~\cite{Marchesano:2010bs}.} and thus we can consistently
take $\psi_{0}^{\pm}$, the superpartner of $A_{\mu}^{\pm}$, to vanish.
When the 4d momentum is zero we have
\begin{subequations}
\label{eq:flat_dirac}
\begin{align}
  0=&
  \bar{\partial}_{\bar{1}}\psi_{1}^{\pm}
  -\bar{\partial}_{\bar{2}}\psi_{2}^{\pm}
  \mp \ui q z^{2}\psi_{3}^{\pm}\label{eq:D-term},\\
  0=&
  {\partial}_{{2}}\psi_{3}^{\pm}
  \mp \ui q \bar{z}^{\bar{2}}\psi_{2}^{\pm},
  \label{eq:F-term1}\\
  0=&
  {\partial}_{{1}}\psi_{3}^{\pm}
  \pm \ui q \bar{z}^{\bar{2}}\psi_{1}^{\pm},
  \label{eq:F-term2}\\
  0=&{\partial}_{{1}}\psi_{2}^{\pm}
  +{\partial}_{{2}}\psi_{1}^{\pm}.  \label{eq:F-term3}
\end{align}
\end{subequations}
The equations~\eqref{eq:flat_dirac} also follow from the conditions
for
supersymmetry~\cite{Cecotti:2009zf,Marchesano:2010bs,Aparicio:2011jx}.
These coupled first-order equations can be turned into largely
decoupled second-order equations by taking derivatives.  For example,
application of $\partial_{1}$ to~\eqref{eq:D-term} and substitution
of~\eqref{eq:F-term2} and~\eqref{eq:F-term3} yields
\begin{subequations}
\begin{equation}
  0=\partial_{1}\bar{\partial}_{\bar{1}}\psi_{1}^{\pm}
  +\partial_{2}\bar{\partial}_{\bar{2}}\psi_{1}^{\pm}
  -q^{2}\left\lvert z^{2}\right\rvert\psi_{1}^{\pm}.
\end{equation}
Similarly,
\begin{align}
  0=&\partial_{1}\bar{\partial}_{\bar{1}}\psi_{2}^{\pm}
  +\partial_{2}\bar{\partial}_{\bar{2}}\psi_{2}^{\pm}
  \pm \ui q \psi_{3}^{\pm}
  -q^{2}\left\lvert z^{2}\right\rvert\psi_{2}^{\pm},\\
  0=&\partial_{1}\bar{\partial}_{\bar{1}}\psi_{3}^{\pm}
  +\partial_{2}\bar{\partial}_{\bar{2}}\psi_{3}^{\pm}
  \mp \ui q \psi_{2}^{\pm}
  -q^{2}\left\lvert z^{2}\right\rvert\psi_{3}^{\pm}.
  \label{eq:2nd_order_eom_2}
\end{align}
\end{subequations}
Using~\eqref{eq:F-term1},~\eqref{eq:2nd_order_eom_2} gives an equation
for $\psi_{3}^{\pm}$ alone.  Writing
$\psi_{3}^{\pm}=\bar{z}^{\bar{2}}\psi_{\pm}$, we have
\begin{equation}
  \label{eq:flat_psi}
  0=\partial_{1}\bar{\partial}_{\bar{1}}\psi_{\pm}
  +\partial_{2}\bar{\partial}_{\bar{2}}\psi_{\pm}
  -q^{2}\left\lvert z^{2}\right\rvert\psi_{\pm}.
\end{equation}
Once $\psi_{\pm}$ is determined, $\psi_{1,2,3}^{\pm}$ are easily found.

\begin{figure}[ht]
  \psfrag{y}{$\sig_{\pm}$}
  \psfrag{r}{$r_{2}/\sqrt{q}$}
  \begin{center}
    \includegraphics[scale=1.05]{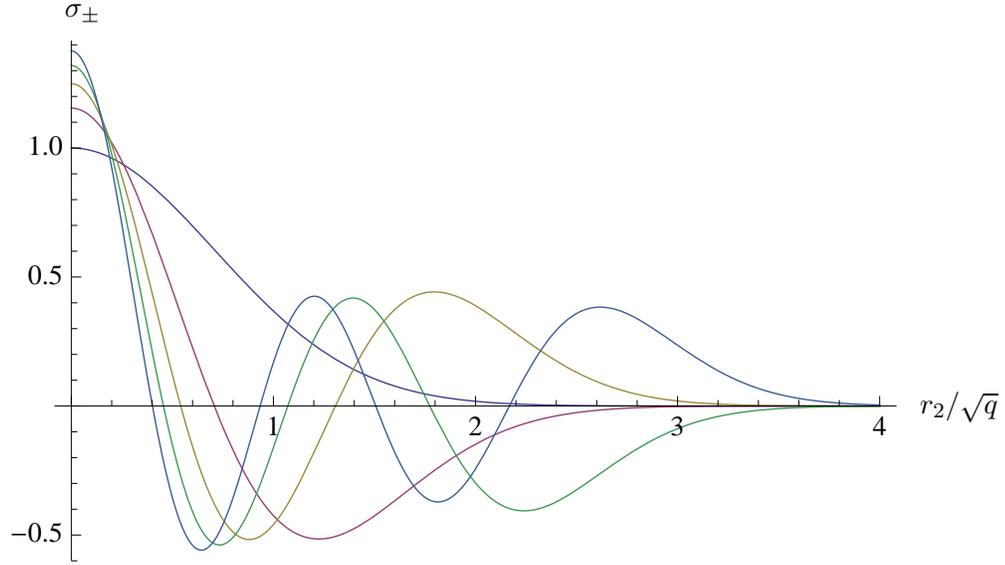}
  \end{center}
  \caption{\label{fig:flat_m2_0}Transverse profiles for the flat space
    vector-like bifundamental modes $\sig_{\pm}$ given
    by~\eqref{eq:flat_solns} for $m_{2}=0$ and $n=0$ (the curve with
    smallest value at $r_{2}=0$) through $n=4$ (the curve with the
    largest value at $r_{2}=0$). The solutions have been normalized to
    the same value using the inner product $\int \ud
    r^{2}f\left(r^{2}\right)g\left(r^{2}\right)$.}
\end{figure}

\begin{figure}[ht]
  \psfrag{y}{$\sig_{\pm}$}
  \psfrag{r}{$r_{2}/\sqrt{q}$}
  \begin{center}
    \includegraphics[scale=1.05]{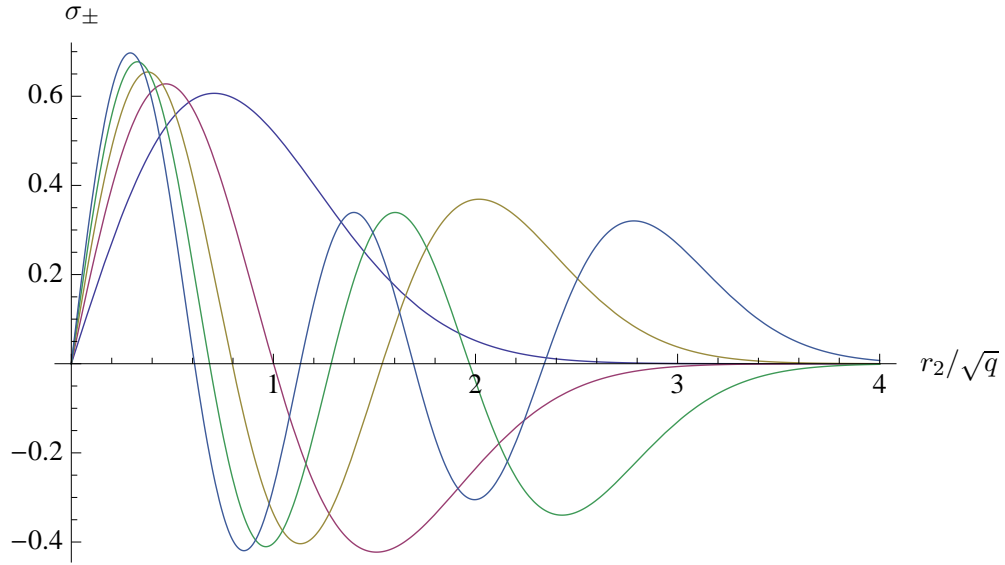}
  \end{center}
  \caption{\label{fig:flat_m2_1}Similar plot as
    figure~\ref{fig:flat_m2_0} except with $m_{2}=1$.}
\end{figure}

Equation~\eqref{eq:flat_psi} is separable.  Performing polar decompositions
$z^{a}=r_{a}\ue^{\ui\phi_{a}}$ and taking the ansatz
\begin{equation}
  \label{eq:flat_ansatz}
  \psi_{\pm}=\ue^{\ui\left(m_{1}\phi_{1}+m_{2}\phi_{2}\right)}
  \zeta_{\pm}\bigl(r_{1}\bigr)\sig_{\pm}\bigl(r_{2}\bigr),
\end{equation}
where $m_{i}$ are integers, we have
\begin{subequations}
\begin{align}
  0=&\zeta''_{\pm}+\frac{1}{r_{1}}\zeta'_{\pm}
  -\frac{m_{1}^{2}}{r_{1}^{2}}\zeta_{\pm}
  -4\lambda\zeta_{\pm},\\
  0=&\sig''_{\pm}-\frac{1}{r_{2}}\sig'_{\pm}
  -\frac{m_{2}^{2}}{r_{2}^{2}}\sig_{\pm}
  -4 q^{2}r_{2}^{2}\sig_{\pm}+4\lambda\sig_{\pm},
\end{align}
\end{subequations}
in which $\lambda$ is a constant to be determined by boundary
conditions.  Imposing that $\sig_{\pm}\to 0$ as $r_{2}\to 0$ we find
\begin{subequations}
\label{eq:flat_solns}
\begin{align}
  \zeta_{\pm}\bigl(r_{1}\bigr)=& c_{1}I_{\left\lvert m_{1}\right\rvert}
  \bigl(\sqrt{2\lambda}\,r_{1}\bigr)
  + c_{2}K_{\left\lvert m_{1}\right\rvert}
  \bigl(\sqrt{2\lambda}\,r_{1}\bigr),\\
  \sig_{\pm}\bigl(r_{2}\bigr)=&\ue^{-qr_{2}^{2}}
  \bigl(2qr_{2}^{2}\bigr)^{\left\lvert m_{2}\right\rvert/2}
  L_{n}^{\left\lvert m_{2}\right\rvert}\bigl(2qr_{2}^{2}\bigr).
\end{align}
\end{subequations}
in which $L_{\nu}^{\mu}$ are the associated Laguerre polynomials,
$I_{\mu}$ and $K_{\mu}$ are the modified Bessel functions of the first
and second kinds, and
\begin{equation}
  \label{eq:6d_spectrum}
  \lambda=q\bigl(2n+\left\lvert m_{2}\right\rvert+1\bigr).
\end{equation}
Regularity of $\sig_{\pm}$ requires that $n$ is a non-negative
integer.  Some of these modes are plotted in
figures~\ref{fig:flat_m2_0} and~\ref{fig:flat_m2_1}.

One may notice that the system~\eqref{eq:flat_dirac} also admits a
zero mode that depends only on $r_{2}$,
\begin{equation}
  \psi_{3}^{\pm}=\ue^{-q r_{2}^{2}}.
\end{equation}
It is easy to confirm that this gives a solution
to~\eqref{eq:flat_psi}, but this solution is not normalizable with
respect to the norm defined by treating~\eqref{eq:flat_psi} as a
Sturm-Liouville problem.  This is a consequence of the fact that the
bifundamental modes are more properly encoded by linear combinations
of the $\psi_{m}^{\pm}$ rather than by the $\psi_{m}^{\pm}$
themselves~\cite{Hashimoto:2003xz,*Nagaoka:2003zn}.  Correspondingly,
the measure used in integrating over the $z^{2}$ and
$\bar{z}^{\bar{2}}$ directions is not that defined
by~\eqref{eq:flat_psi} (see~\cite{Marchesano:2010bs}).  However, aside
from this zero mode, the above equations successfully reproduce the
spectrum of 6d masses~\eqref{eq:6d_spectrum}.

\subsection{Chiral modes}  \label{flatspacechiral}

We now consider magnetized intersections since, upon compactification,
such a construction gives a chiral 4d theory. We will again focus on
supersymmetric configurations, which implies that $F_{2}$ must be
$\left(1,1\right)$ and primitive. Consider first a single D7-brane on
$\mathbb{R}^{3,1}\times \mathbb{C}^{2}$.  The most general
$\left(1,1\right)$ flux that can be supported by the D7-brane is
\begin{equation}
  \label{eq:general_flux}
  F_{2}=-\frac{\ui}{2}f_{1}\,\ud z^{1}\wedge\ud\bar{z}^{\bar{1}}
  -\frac{\ui}{2}f_{2}\,\ud z^{2}\wedge\ud\bar{z}^{\bar{2}}
  -\frac{\ui}{2}g_{1}\,\ud z^{1}\wedge\ud\bar{z}^{\bar{2}}
  -\frac{\ui}{2}g_{2}\,\ud \bar{z}^{\bar{1}}\wedge\ud{z}^{{2}}.
\end{equation}
The $1\bar{1}$ component
will describe
the magnetization of the intersection, and so we will look for the
simplest configurations with $f_{1}\neq 0$.  The K\"ahler form on  
$\mathbb{C}^{2}$ is simply
\begin{equation}
  J=-\frac{\ui}{2}\sum_{I=1}^{2}\ud z^{I}\wedge\ud\bar{z}^{\bar{I}},
\end{equation}
and so primitivity imposes $f_{1}=-f_{2}$.
The Bianchi identity implies that $f_{1}$ is harmonic,
\begin{equation} \label{bianchi}
  0=\partial_{1}\bar{\partial}_{1}f_{1}+
  \partial_{2}\bar{\partial}_{2}f_{1}.
\end{equation}
In the absence of sources,~\eqref{bianchi} requires that $f_{1}$ is constant.  We
can then consistently set $g_{1}=g_{2}=0$ and obtain the supersymmetric
magnetization
\begin{equation}
  \label{eq:mag_flux}
  F_{2}=-\ui\,M\,\bigl\{\ud z^{1}\wedge\ud\bar{z}^{\bar{1}}
  -\ud z^{2}\wedge\ud\bar{z}^{\bar{2}}\bigr\}.
\end{equation}
Compactification would impose a quantization condition on $M$, but in
the non-compact case we can freely take $M$ to be any constant.  The
above magnetization follows from the gauge choice
\begin{equation}
  \label{eq:mag_connection}
  A_{1}=-\frac{\ui}{2}M\bigl\{z^{1}\ud\bar{z}^{\bar{1}}-
  \bar{z}^{\bar{1}}\ud z^{1}
  -z^{2}\ud\bar{z}^{\bar{2}}
  +\bar{z}^{\bar{2}}\ud z^{2}\bigr\}.
\end{equation}

To obtain chiral matter, we again consider the intersection of two
D7-branes described by the Higgsing~\eqref{eq:Higgs}, and choose a
magnetization
\begin{equation}
  F_{2}=-M\begin{pmatrix}1 & \\ & -1\end{pmatrix}
  \,\bigl\{\ud z^{1}\wedge\ud\bar{z}^{\bar{1}}
  -\ud z^{2}\wedge\ud\bar{z}^{\bar{2}}\bigr\}.   \label{magnetizationchoice}
\end{equation}
The corresponding connection is
\begin{equation}
  A_{1}=-\frac{\ui}{2}M
  \begin{pmatrix}1 & \\ & -1\end{pmatrix}
  \bigl\{z^{1}\ud\bar{z}^{\bar{1}}-
  \bar{z}^{\bar{1}}\ud z^{1}
  -z^{2}\ud\bar{z}^{\bar{2}}
  +\bar{z}^{\bar{2}}\ud z^{2}\bigr\}.
\end{equation}
For simplicity of presentation we will take $M>0$.

With a non-trivial connection, the equation of motion for the fermions
becomes
\begin{equation}
  0=\hat{\Ga}^{\alpha}D_{\alpha}\theta
  -\ui\,\hat{\Ga}_{i}\bigl[\Phi^{i},\theta\bigr],
\end{equation}
where $D_{\alpha}$ is the gauge-covariant derivative.  Following the
same decomposition and procedure as for the vector-like case, we again
find~\eqref{eq:flat_dirac} up to the replacements
\begin{align}
  \partial_{1}\psi_{m}^{\pm}&\to
  \bigl(\partial_{1}\pm M\bar{z}^{\bar{1}}\bigr)\psi_{m}^{\pm},
  &\bar{\partial}_{\bar{1}}\psi_{m}^{\pm}&\to
  \bigl(\bar{\partial}_{\bar{1}}\mp M{z}^{{1}}\bigr)\psi_{m}^{\pm},\notag\\
  \partial_{2}\psi_{m}^{\pm}&\to
  \bigl(\partial_{2}\mp M\bar{z}^{\bar{2}}\bigr)\psi_{m}^{\pm},
  &\bar{\partial}_{\bar{2}}\psi_{m}^{\pm}&\to
  \bigl(\bar{\partial}_{\bar{2}}\pm M{z}^{{2}}\bigr)\psi_{m}^{\pm}.
\end{align}
Again writing $\psi_{3}^{\pm}=\bar{z}^{\bar{2}}\psi_{\pm}$, we find
\begin{equation}
  \label{eq:magnetized_eom}
  0=\biggl\{\partial_{1}\bar{\partial}_{\bar{1}}+
  \partial_{2}\bar{\partial}_{\bar{2}}
  \pm M\bigl(\bar{z}^{\bar{1}}\bar{\partial}_{\bar{1}}
  -z^{1}\partial_{1}
  -\bar{z}^{2}\bar{\partial}_{\bar{2}}
  +z^{2}\partial_{2}\bigr)
  -M^{2}\bigl\lvert z^{1}\bigr\rvert^{2}
  -\bigl(M^{2}+q^{2}\bigr)\bigl\lvert z^{2}\bigr\rvert^{2}
  \biggr\}\psi_{\pm}.
\end{equation}

Due to the self-duality of the magnetic
flux,~\eqref{eq:magnetized_eom} is separable.
Again using the polar decomposition $z^{a}=r_{a}\ue^{\ui\phi_{a}}$ and
taking the ansatz~\eqref{eq:flat_ansatz}, we find the equations
\begin{align}
  0=&\zeta_{\pm}''+\frac{1}{r_{1}}\zeta_{\pm}'
  -\frac{m_{1}^{2}}{r_{1}^{2}}\zeta_{\pm}
  -4M^{2}r_{1}^{2}\zeta_{\pm}
  +\bigl(-4\lambda\pm 4 M m_{1}\bigr)\zeta_{\pm},\\
  0=&\sig_{\pm}''+\frac{1}{r_{2}}\sig_{\pm}'
  -\frac{m_{2}^{2}}{r_{2}^{2}}\sig_{\pm}
  -4\ka^{2}r_{2}^{2}\sig_{\pm}
  +\bigl(4\lambda\mp 4Mm_{2}\bigr)\sig_{\pm},
\end{align}
in which
\begin{equation}
  \ka=\sqrt{M^{2}+q^{2}},
\end{equation}
and $\lambda$ is again a constant to be determined by boundary
conditions.   The solutions are
\begin{align}  \label{conflusols}
  \zeta_{\pm}\bigl(r_{1}\bigr)=&\ue^{-M r_{1}^{2}}
  \bigl(2Mr_{1}^{2}\bigr)^{\left\lvert m_{1}\right\rvert/2}\biggl\{
  \cM\bigl(\alpha;m_{1}+1;2Mr_{1}^{2}\bigr)+
  \cU\bigl(\alpha;m_{1}+1;2Mr_{1}^{2}\bigr)\biggr\}\notag\\
  \sig_{\pm}\bigl(r_{2}\bigr)=&\ue^{-\ka r_{2}^{2}}
  \bigl(2\ka r_{2}^{2}\bigr)^{\left\lvert m_{2}\right\rvert/2}
  L_{n_{2}}^{\left\lvert m_{2}\right\rvert}
  \bigl(2\ka r_{2}^{2}\bigr),
\end{align}
with
\begin{equation}
  \lambda=\ka\bigl(2n_{2}+\left\lvert m_{2}\right\rvert+1\bigr)
  \pm M m_{2}
  \equiv M\bigl(2\alpha-\bigl\lvert m_{1}\bigr\rvert\pm m_{1}-1\bigr),
\end{equation}
where the final relation defines $\alpha$.  In~\eqref{conflusols},
$\cM$ and $\cU$ are the confluent hypergeometric functions of the
first and second kinds,\footnote{Since the confluent hypergeometric
  function $\phantom{}_{1}F_{1}\left(a;b;z\right)$ is not defined when
  $b=0,-1,-2,\ldots$, we use the regularized version
  $\cM\left(a;b;z\right)=\phantom{}_{1}F_{1}\left(a;b;z\right)/\Ga\left(b\right)$.}
and regularity requires that $n_2$ be a non-negative integer.

The chirality of the spectrum is a consequence of the different
behavior of the different charges.  It is most easily seen by
considering the ``missing'' zero
mode~\cite{Cremades:2004wa,Marchesano:2010bs}
\begin{equation}
  \psi_{3}^{\pm}\sim \ue^{-\ka r_{2}^{2}}\ue^{\mp M r_{1}^{2}}
  h\bigl(z^{1}\bigr),
\end{equation}
where $h$ is a holomorphic function of $z^{1}$.  Since we have taken
$M>0$, only the $+$ sector gives rise to normalizable modes, and hence
the spectrum is chiral.  The fact that $h$ is an arbitrary holomorphic
function indicates that there are an infinite number of such chiral
modes, as is consistent with the fact that the chiral index, which is
proportional to $\int F_{2}$, is divergent.  Upon compactification,
further conditions are imposed on $h\bigl(z^{1}\bigr)$ (see
e.g.~\cite{Cremades:2004wa}) and the spectrum becomes finite.

%%%%
%%%%
\section{\label{sec:vec_mesons}Non-chiral Mesons from D7-branes in AdS}
%%%%

We now consider vector-like mesons arising on intersecting D7-branes
in $AdS_{5}\times S^{5}$, building on the groundwork laid in
\S\ref{sec:unwarped}. As discussed in the introduction, the
configuration of interest is the gravity dual of $\cN=4$ $\SU{N}$ SYM
with a $\U{1}\times\U{1}$ flavor group.  The strings stretching
between the D7-branes are dual to mesonic operators with charges
$\left(\pm 1, \mp 1\right)$ under this $\U{1}\times\U{1}$. Our
analysis has much in common with the treatment of intersecting
D7-branes in weakly warped geometries~\cite{Marchesano:2010bs};
however, $AdS_{5}\times S^{5}$ is strongly warped in the sense that no
limit of the geometry reproduces a factorized geometry
$\RR^{3,1}\times X^{6}$, and so we will need to use different
techniques to solve the resulting equations of motion.

\subsection{Setup and equations of motion}

The metric for $AdS_{5}\times S^{5}$ can be written as a warped
product of $\mathbb{R}^{3,1}$ and $\mathbb{C}^{3}$,
\begin{equation}
  \ud s_{10}^{2}=\ue^{2\cA}\eta_{\mu\nu}\ud x^{\mu}\ud x^{\nu}
  +\ue^{-2\cA}\ud z^{I}\ud\bar{z}^{\bar{I}},\qquad
  \cA=\frac{1}{2}\log\frac{z^{I}\bar{z}^{\bar{I}}}{L^{2}}.
\end{equation}
Using hyperspherical coordinates on $\mathbb{C}^{3}= \mathbb{R}^{6}$,
this becomes the familiar metric for $AdS_{5}\times S^{5}$,
\begin{equation}
  \ud s_{10}^{2}=\frac{R^{2}}{L^{2}}\eta_{\mu\nu}\ud x^{\mu}\ud x^{\nu}
  +\frac{L^{2}}{R^{2}}\ud R^{2}+L^{2}\ud s_{S^{5}}^{2},
\end{equation}
where $\ud s_{S^{5}}^{2}$ is the standard metric on a unit
$S^{5}$. The geometry is supported by the $5$-form flux
\begin{equation}
  F_{5}=\bigl(1+\hat{\ast}\bigr)g_{\us}^{-1}\ud\ue^{4\cA}\wedge
  \ud\vol_{\mathbb{R}^{3,1}},
\end{equation}
where $\hat{\ast}$ is the 10d Hodge  star.  In the presence of such
flux, the action for a single D7-brane
becomes~\cite{Marolf:2003ye,*Marolf:2003vf,*Martucci:2005rb}
\begin{multline}
  \label{eq:warped_abelian}
  S_{\uD 7}=-\frac{1}{g_{8}^{2}}\int_{\cW}\ud^{8}\xi^{\alpha}\sqrt{-\hat{g}}\,
  \biggl\{\frac{1}{2}\hat{g}_{ij}\hat{g}^{\alpha\beta}
  \partial_{\alpha}\Phi^{i}\partial_{\beta}\Phi^{j}
  +\frac{1}{4}\hat{g}^{\alpha\beta}\hat{g}^{\ga\delta}
  F_{\alpha\ga}F_{\beta\delta}+
  \ui\bar{\Theta}P_{-}^{\uD 7}
  \hat{g}^{\alpha\beta}\hat{\Ga}_{\alpha}
  \hat{\nabla}_{\beta}\Theta \\
  +\frac{g_{\us}}{8\cdot 4!}
  \hat{\ep}^{\alpha_{1}\cdots\alpha_{8}}C_{\alpha_{1}\cdots\alpha_{4}}
  F_{\alpha_{5}\alpha_{6}}F_{\alpha_{7}\alpha_{8}}
  +\frac{\ui g_{\us}}{16}\bar{\Theta}
  P_{-}^{\uD 7}
  \hat{g}^{\alpha\beta}\hat{\Ga}_{\alpha}
  \hat{\slashed{F}}_{5}\hat{\Ga}_{\beta}
  \bigl(\ui\sig_{2}\bigr)
  \Theta\biggr\},
\end{multline}
in which
\begin{equation}
  \hat{\slashed{F}}_{5}
  =\frac{1}{5!}
  F_{M_{1}\cdots M_{5}}\hat{\Ga}^{M_{1}\cdots M_{5}},
\end{equation}
is constructed by contracting all indices of $F_{5}$ with
$\hat{\Ga}$-matrices, and not just those along the worldvolume.  If the
D7-brane fills $\mathbb{R}^{3,1}$ and a cycle $\cS^{4}$ in the other
directions, then after $\ka$-fixing to~\eqref{eq:kappa-gauge} and taking
into account the nontrivial spin connection, the fermionic
contribution to the action is~\cite{Marchesano:2008rg}
\begin{equation}
  S_{\uD 7}^{\uF}=-\frac{\ui}{2g_{8}^{2}}
  \int\ud^{8}\xi^{\alpha}\sqrt{-\hat{g}}\,
  \bar{\theta}
  \biggl\{\hat{g}^{\alpha\beta}\hat{\Ga}_{\alpha}\partial_{\beta}
  +\frac{1}{2}\hat{g}^{\alpha\beta}\hat{\Ga}_{\alpha}\partial_{\beta}\cA
  \bigl(1+2\hat{\Ga}_{\cS^{4}}\bigr)\biggr\}\theta,
\end{equation}
in which
\begin{equation}
  \hat{\Ga}_{\cS^{4}}=\ud\hat{\slashed{\vol}}_{\cS^{4}},
\end{equation}
is the chirality operator on $\cS^{4}$.

In the non-Abelian case, closed-string fields like the warp factor are
interpreted as Taylor series in the adjoint-valued transverse
deformations, and thus the closed-string fields are themselves
adjoint-valued~\cite{Myers:1999ps}.  However, as in the unwarped case,
this fact is only important for higher-dimension operators, and can
be neglected to leading order in
$\ell_{\us}$.
Similar terms are expected
in the non-Abelian fermionic action, but have not been computed explicitly.
 However,
to leading order
in $\ell_{\us}$, supersymmetry and gauge-invariance require that the
action take the form~\cite{Wynants,Marchesano:2010bs}
\begin{equation}
  S_{\uD 7}^{\uF}=-\frac{\ui}{2g_{8}^{2}}
  \int\ud^{8}\xi^{\alpha}\sqrt{-\hat{g}}\,\tr\biggl\{
  \bar{\theta}\,
  \hat{\Ga}^{\alpha}D_{\alpha}\theta
  +\frac{1}{2}\bar{\theta}\,
  \hat{\Ga}^{\alpha}\partial_{\alpha}\cA
  \bigl(1+2\hat{\Ga}_{\cS^{4}}\bigr)\theta
  -\frac{1}{2}\bar{\theta}\,\hat{\Ga}_{i}  \bigl[\Phi^{i},\theta\bigr]\biggr\}.
\end{equation}

The intersection of two D7-branes satisfying
\begin{equation}
  \label{eq:intersecting}
  \uD 7_{1}: z^{3}=\mu+tz^{2},\quad
  \uD 7_{2}: z^{3}=\mu-tz^{2},
\end{equation}
is described by
\begin{equation}
  \label{eq:vev2}
  \Phi=\begin{pmatrix}
    \ell_{\us}^{-2}\mu + qz^{2} & \\
    & \ell_{\us}^{-2}\mu-qz^{2}.
  \end{pmatrix}.
\end{equation}
When $\mu=0$, the D7-branes reach the origin of warping and the dual
quarks are massless: in the $\uD$-brane picture, the D3-branes and
D7-branes intersect and the strings stretching between them have zero
length.  However, when there is a finite separation between the
branes, the quarks have a mass proportional to $\mu$.  Consequently,
the mesonic spectrum becomes gapped~\cite{Kruczenski:2003be}. The warp
factor is to be evaluated at this vev, but so long as $t$ is
sufficiently small, on the D7-brane we can take
\begin{equation}
  \cA=\frac{1}{2}\log
  \frac{\left\lvert z^{1}\right\rvert^{2}
    +\left\lvert z^{1}\right\rvert^{2}+\mu^{2}}{L^{2}}.
\end{equation}
Decomposing $\theta$ as~\eqref{eq:chirality_split} and matching terms
of internal chirality, we find
\begin{subequations}
\begin{align}
  0=&
  \bigl(\bar{\partial}_{\bar{1}}-\frac{1}{2}\bar{\partial}_{\bar{1}}\cA\bigr)
  \psi_{1}^{\pm}
  -\bigl(\bar{\partial}_{\bar{2}}-\frac{1}{2}\bar{\partial}_{\bar{2}}\cA\bigr)
  \psi_{2}^{\pm}
  \mp \ui q \ue^{-2\cA}z^{2}\psi_{3}^{\pm},\\
  0=&
  \bigl({\partial}_{{2}}+\frac{3}{2}\partial_{2}\cA\bigr)\psi_{3}^{\pm}
  \mp \ui q\ue^{-2A} \bar{z}^{\bar{2}}\psi_{2}^{\pm},\\
  0=&
  \bigl({\partial}_{{1}}+\frac{3}{2}\partial_{1}\cA\bigr)\psi_{3}^{\pm}
  \pm \ui q \ue^{-2A}\bar{z}^{\bar{2}}\psi_{1}^{\pm},\\
  0=&\bigl({\partial}_{{1}}-\frac{1}{2}\partial_{1}\cA\bigr)\psi_{2}^{\pm}
  +\bigl({\partial}_{{2}}-\frac{1}{2}\partial_{2}\cA\bigr)\psi_{1}^{\pm},
\end{align}
\end{subequations}
where, as in the flat space analysis of \S\ref{sec:unwarped}, we
have evaluated the equations at zero 4d momentum and have set
$\psi_{0}^{\pm}=0$.  Taking, as in~\cite{Marchesano:2008rg}
\begin{equation}
  \label{eq:warp_factor_factor}
  \psi_{1,2}^{\pm}=\ue^{\cA/2}\vp_{1,2}^{\pm},\quad
  \psi_{3}^{\pm}=\ue^{-3\cA/2}\vp_{3}^{\pm},
\end{equation}
and finally writing $\vp_{3}^{\pm}=\bar{z}^{\bar{2}}\vp_{\pm}$, we
find the warped analogue of~\eqref{eq:flat_psi}
\begin{equation}
  \label{eq:warped_psi}
  0=\bigl\{\partial_{1}\bar{\partial}_{\bar{1}}+
  \partial_{2}\bar{\partial}_{\bar{2}}
  -q^{2}\left\lvert z^{2}\right\rvert^{2}\ue^{-4\cA}\bigr\}\vp_{\pm}.
\end{equation}

Since the warp factor depends on both $z^{1}$ and $z^{2}$,
(\ref{eq:warped_psi})
is not
separable in those variables.  However, writing
\begin{equation}
  \label{eq:beta_coords}
  z^{1}=r\cos\beta\,\ue^{\ui\phi_{1}},\quad
  z^{2}=r\sin\beta\,\ue^{\ui\phi_{2}},
\end{equation}
equation \eqref{eq:warped_psi} becomes
\begin{equation}
  \label{eq:warped_psi_2}
  0=\biggl\{
  \partial_{r}^{2}
  +\frac{3}{r}\partial_{r}
  +\frac{1}{r^{2}}
  \breve{\nabla}^{2}
  -\frac{4q^{2}r^{2}L^{4}\sin^{2}\beta}{\left(r^{2}+\mu^{2}\right)^{2}}
  \biggr\}\vp_{\pm},
\end{equation}
in which
\begin{equation}
  \breve{\nabla}^{2}=
  \partial_{\beta}^{2}
  +\bigl(\cot\beta-\tan\beta\bigr)\partial_{\beta}
  +\frac{1}{\cos^{2}\beta}\partial_{\phi_{1}}^{2}
  +\frac{1}{\sin^{2}\beta}\partial_{\phi_{2}}^{2}
\end{equation}
is the Laplacian on a unit $S^{3}$ (see Appendix~\ref{app:spherical}).

When $\mu=0$,~\eqref{eq:warped_psi_2} is completely separable.
Indeed, taking
\begin{equation}
  \label{eq:split_ansatz}
  \vp_{\pm}=\ue^{\ui\left(m_{1}\phi_{1}+m_{2}\phi_{2}\right)}
  f_{\pm}\bigl(r\bigr)Q_{\pm}\bigl(\cos2\beta\bigr),
\end{equation}
we find that the radial equation satisfies
\begin{equation}
  0=f_{\pm}''+\frac{3}{r}f_{\pm}'-\frac{\lambda}{r^{2}}f_{\pm},
  \label{eq:vector_radial}
\end{equation}
while the $\beta$ equation is
\begin{equation}
  0=4\bigl(1-x^{2}\bigr)Q_{\pm}''-8xQ_{\pm}'-\frac{2m_{1}^{2}}{1+x}Q_{\pm}
  -\frac{2m_{2}^{2}}{1-x}Q_{\pm}-2\xi^{2}\bigl(1-x\bigr)Q_{\pm}
  +\lambda Q_{\pm},
  \label{eq:angular}
\end{equation}
in which $x=\cos 2\beta$,
\begin{equation}  \label{xidef}
\xi^{2} \equiv q^{2}L^{4} = \frac{1}{\pi} t^2 g_{\us} N ,
\end{equation}
and $\lambda$ is a constant to be determined by boundary
conditions.\footnote{Note that in this section and the next, $\lambda$
  carries no dimensions, in contrast to the previous section.}

\subsection{The meson spectrum}

When $\xi=0$, the solutions to~\eqref{eq:angular} are the scalar
hyperspherical harmonics (see Appendix~\ref{app:spherical})
\begin{equation}
  \label{eq:spherical_harm}
  Q_{\pm}\bigl(x\bigr)
  =c
  \bigl(1+x\bigr)^{m1/2}\bigl(1-x\bigr)^{m2/2}
  P_{\frac{1}{2}\left(\ell-m_{1}-m_{2}\right)}^{\left(m_{2},m_{1}\right)}\bigl(x\bigr),
\end{equation}
where $P_{n}^{\left(a,b\right)}$ are the Jacobi Polynomials, $c$ is
the normalization constant~\eqref{eq:norm},
$\lambda=\ell\left(\ell+2\right)$, and the quantum numbers must
satisfy the inequalities $0\le\left \lvert
  m_{1}\right\rvert+\left\lvert m_{2}\right\rvert\le \ell$ and the
constraint $\frac{1}{2}\left(\ell-m_{1}-m_{2}\right)\in \ZZ$.

We have been unable to find analytic solutions to~\eqref{eq:angular}
when $\xi\neq 0$.  However, since~\eqref{eq:angular} is an ordinary
differential equation, numerical methods readily apply.  We implement
a spectral method by expanding the unknown solution in terms of the
spherical harmonics.  The potential term proportional to $\xi$ does
not mix modes of different $m_{1}$ and $m_{2}$, so we can accomplish
the spectral decomposition by writing
\begin{equation}
  Q\bigl(x\bigr)=\sum_{\ell}b_{\ell}y_{\ell}\bigl(x\bigr),
\end{equation}
where $y_{\ell}$ are the solutions~\eqref{eq:spherical_harm} and we have
suppressed other indices.  Equation~\eqref{eq:angular} then becomes
\begin{equation}
  \label{eq:angular_2}
  0=\sum_{\ell}\biggl\{\lambda-\ell\left(\ell+2\right)
  -2\xi^{2}\bigl(1-x\bigr)\biggr\}b_{\ell}y_{\ell}.
\end{equation}
Using that at fixed $m_{1}$ and $m_{2}$,
\begin{equation}
  \int_{-1}^{1}\ud x\, y_{\ell}y_{\ell'}=\frac{1}{\pi^{2}}\delta_{\ell\ell'},
\end{equation}
and using the recursion relationship~\eqref{eq:recursion}, we can
re-express~\eqref{eq:angular_2} as the matrix equation
\begin{equation}
  \label{eq:spectral}
  0=\biggl[\lambda-\ell\left(\ell+2\right)-2\xi^{2}d_{0}\biggr]b_{\ell}
  +2\xi^{2}d_{-}b_{\ell-2}+2\xi^{2}d_{+}b_{\ell+2},
\end{equation}
with
\begin{align}
  d_{0}=&\biggl(1+
  \frac{m_{2}^{2}-m_{1}^{2}}
  {\ell\left(\ell+2\right)}
  \biggr),\notag\\
  d_{-}=&\frac{1}{2\ell}
  \sqrt{\frac{\left(\ell+m_{1}+m_{2}\right)\left(\ell-m_{1}-m_{2}\right)
      \left(\ell+m_{1}-m_{2}\right)\left(\ell-m_{1}+m_{2}\right)}
    {\ell^{2}-1}},\\
  d_{+}=&\frac{1}{2\left(\ell+2\right)}
  \sqrt{\frac{\left(\ell+2+m_{1}+m_{2}\right)\left(\ell+2-m_{1}-m_{2}\right)
      \left(\ell+2+m_{1}-m_{2}\right)\left(\ell+2-m_{1}+m_{2}\right)}
    {\left(\ell+1\right)\left(\ell+3\right)}}.\notag
\end{align}
Note that even and odd $\ell$s do not mix, so that this effectively
gives two independent matrix equations where the matrices are each
tridiagonal.

\begin{figure}
  \psfrag{x}{$\xi$}
  \psfrag{l}{$\lambda$}
  \begin{center}
    \includegraphics[scale=1]{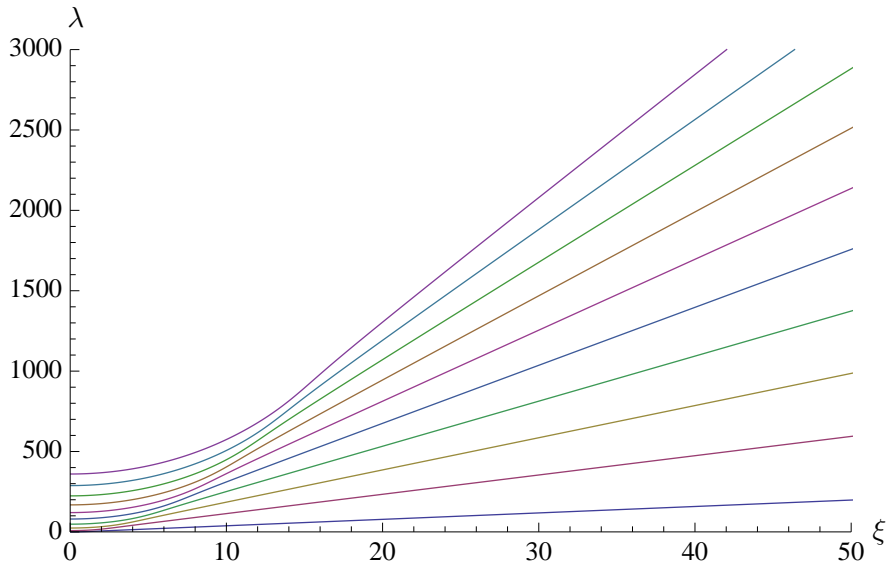}
  \end{center}
  \caption{\label{fig:angular_spec_m1_0_m2_0} The first few eigenvalues
    of~\eqref{eq:angular} found via spectral methods, for
    $m_{1}=m_{2}=0$.  The growth continues to be linear as $\xi$
    increases.}
\end{figure}

\begin{figure}
  \psfrag{n}{$\ell$}
  \psfrag{l}{$\lambda$}
  \begin{center}
    \includegraphics[scale=1]{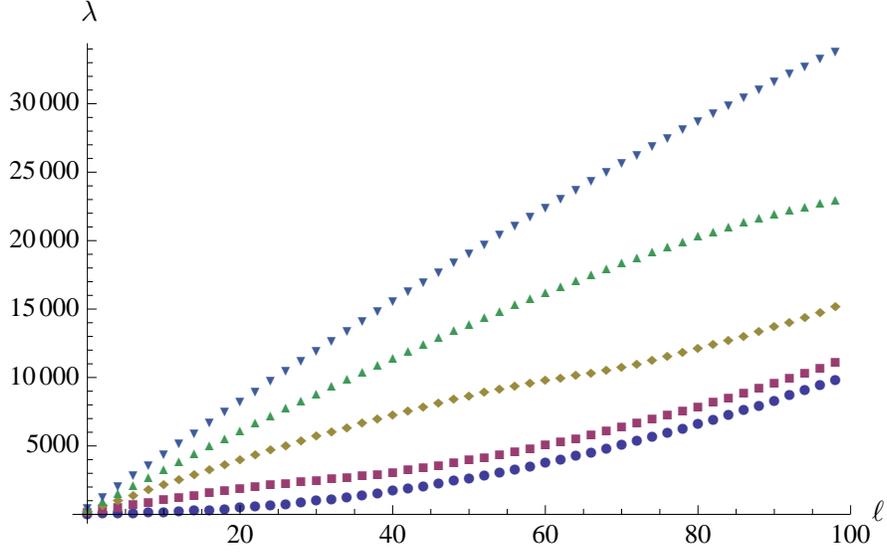}
  \end{center}
  \caption{\label{fig:angular_spec_m1_0_m2_0_2}The spectrum for
    $m_{1}=m_{2}=0$ (which requires that $\ell$ be even) for $\xi=0$
    (bottom), $25$, $50$, $75$, and $100$ (top).}
\end{figure}

\begin{figure}
  \psfrag{b}{$\beta/\pi$}
  \psfrag{y}{$Q$}
  \begin{center}
    \includegraphics[scale=1]{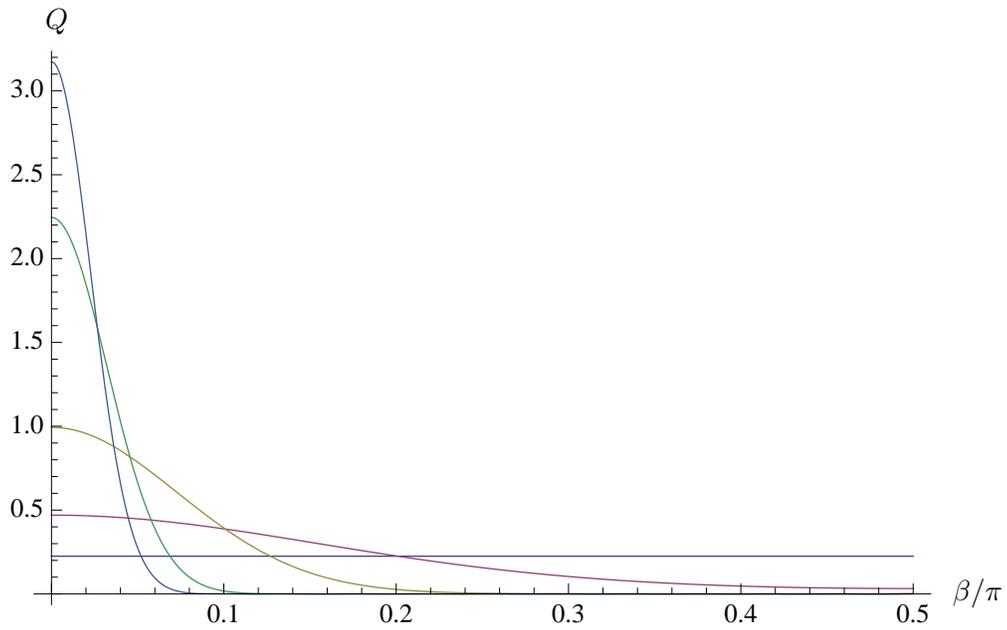}
  \end{center}
  \caption{\label{fig:angular_wavefns_m1_0_m2_0}The lowest-lying solutions
    of~\eqref{eq:angular} for $\xi=0,2.5,10,50,100$.  When $\xi=0$,
    the solution is a constant zero mode, but as $\xi$ increases, the profile
    becomes increasingly peaked at $\beta=0$, the location of the
    intersection.}
\end{figure}

Solving~\eqref{eq:angular} amounts to diagonalization of the matrix
defined by~\eqref{eq:spectral}.  Unfortunately, because this is an
infinite-dimensional matrix, we cannot perform this diagonalization
exactly.  However, to obtain an estimate of the spectrum, we can
truncate the matrix to a finite submatrix.  A good rule of thumb in
such problems is that including the first $2n$ modes determines the
first $n$ eigenvalues to an accuracy of a few percent
\cite{Boyd:2001aa}. Accurate eigenvalues will be robust against
variations in $n$, and our strategy will be to increase the number of
modes included until the eigenvalues calculated in this way stabilize.
The first few eigenvalues at $m_{1}=m_{2}=0$ resulting from this
process are shown in figures~\ref{fig:angular_spec_m1_0_m2_0}
and~\ref{fig:angular_spec_m1_0_m2_0_2}.  As $\xi$ increases, the
wavefunctions become increasingly localized on the intersection at
$\beta=0$, as shown in figure~\ref{fig:angular_wavefns_m1_0_m2_0}.

Note that when $\xi\ll 1$,~\eqref{eq:spectral} immediately yields the
perturbative result
\begin{equation}
  \lambda\approx \ell\left(\ell+2\right)+
  2\xi^{2}\biggl[1+\frac{m_{2}^{2}-m_{1}^{2}}{\ell\left(\ell+2\right)}\biggr].
\end{equation}
However, since $\xi^{2}= t^{2}g_{\us}N/\pi$,
 working at $\xi\ll 1$
requires taking
$t^{2}$ to be small with respect to the inverse 't~Hooft coupling $1/\lambda$.
This limit is of little utility in the present investigation, because we are interested in taking
$\lambda \to \infty$ to suppress $\alpha^{\prime}$ corrections to the leading-order supergravity, cf.~(\ref{decouplinglimit}).

If instead $\xi\gg 1$, we find that the
spectrum is well-approximated by
\begin{equation}
  \lambda\approx 4\xi\bigl(\ell+\left\lvert m_{1}\right\rvert-1\bigr).
\end{equation}
At large $\xi$, $\ell$ is no longer a good quantum number, as the
intersection badly breaks the rotational symmetry of the
$S^{3}$. Correspondingly, the solutions to~\eqref{eq:angular} are
linear combinations of many different spherical harmonics.  However,
$m_{1}$ and $m_{2}$ remain good quantum numbers, and so we find it
more natural to write the spectrum as
\begin{equation}
  \lambda\approx 4\xi\bigl(n+\left\lvert m_{2}\right\rvert+1\bigr),
\end{equation}
where $n=\ell-\left\lvert m_{1}\right\rvert-\left\lvert
  m_{2}\right\rvert$.

With the eigenvalues of~\eqref{eq:angular} in hand, the solution
to~\eqref{eq:vector_radial} is
\begin{equation}
  \label{eq:vector_soln}
  f_{\pm}=c_{1}r^{-1-\sqrt{1+\lambda}}+c_{2}r^{-1+\sqrt{1+\lambda}}.
\end{equation}
We can compare  the solution (\ref{eq:vector_soln})
to the well-known result for a canonically
normalized scalar at zero momentum,
\begin{equation}
  \label{eq:scalar_soln}
  \vp=\vp_{0}r^{\Delta-4}+\vp_{1}r^{-\Delta}.
\end{equation}
The solution~\eqref{eq:vector_soln} does not match the form
(\ref{eq:scalar_soln}), since the transverse deformations are not
canonically normalized (see~\eqref{eq:5d_transverse_theory}).
Nevertheless, $\Delta$ can be determined by taking
the ratio
 of the two
terms in~\eqref{eq:vector_soln}, and we find the result
\begin{equation}
  \Delta=2+\sqrt{1+\lambda}.
\end{equation}
This then gives the approximate expressions
\begin{equation}
  \label{eq:spectrum}
  \Delta\approx
  \begin{cases}
    \ell+3+\frac{\xi^{2}}{1+\ell}
    \left[1+\frac{m_{2}^{2}-m_{1}^{2}}{\ell\left(\ell+2\right)}\right] &
    \xi \ll 1,\\
    2\sqrt{\xi\left(n+\left\lvert m_{2}\right\rvert+1\right)} & \xi \gg 1
  \end{cases}.
\end{equation}

The fact that the radial modes are simply power laws is an indication
that the dual theory is conformal.  Indeed, one can confirm that the
$\mu=0$ configuration~\eqref{eq:intersecting} respects the
supersymmetry generated by eight supercharges, four of which
correspond to the generators of superconformal transformations in the
dual theory.  Alternatively, when $\mu=0$, the vev~\eqref{eq:vev2}
corresponds to a strictly marginal deformation of the theory.  To see
this, it suffices to consider the Abelian
action~\eqref{eq:warped_abelian} and examine only the action of the
transverse scalars $\Phi^{i}$.  Using the complexified field $\Phi$
and expanding in scalar spherical harmonics gives the 5d action
\begin{equation}
  \label{eq:5d_transverse_theory}
  S\sim-\int\ud^{5}x\,\sqrt{-g}\sum_{\ell=0}^{\infty}
  \biggl\{\frac{L^{2}}{r^{2}}
  g^{mn}\partial_{m}\Phi_{\ell}^{\dagger}\partial_{n}\Phi_{\ell}
  +\frac{\ell\left(\ell+2\right)}{r^{2}}\Phi_{\ell}^{\dagger}\Phi_{\ell}
  \biggr\}.
\end{equation}
Defining the canonically normalized scalars
$\chi_{\ell}=\frac{L}{r}\Phi_{\ell}$ gives
\begin{equation}
  S\sim-\int\ud^{5}x\,\sqrt{-g}\sum_{\ell=0}^{\infty}
  \biggl\{
  g^{mn}\partial_{m}\chi_{\ell}^{\dagger}\partial_{n}\chi_{\ell}
  +\frac{\ell\left(\ell+2\right)-3}{L^{2}}\chi_{\ell}^{\dagger}\chi_{\ell}
  \biggr\}.
\end{equation}
Using the familiar result
\begin{equation}
  \Delta=2+\sqrt{4+m^{2}L^{2}}
\end{equation}
yields
\begin{equation}
  \Delta=\ell+3.
\end{equation}
With the coordinates of~\eqref{eq:beta_coords}, the
configuration $\Phi=q z^{2}$ can be expressed as
\begin{equation} \label{deltafour}
  \Phi=q r\,\sin\beta\,\ue^{\ui\phi_{2}}
  =\frac{qr}{\sqrt{2}}\sqrt{1-\cos 2\beta}\,\ue^{\ui\phi_{2}}.
\end{equation}
Comparing to~\eqref{eq:spherical_harm}, the mode~\eqref{deltafour}
corresponds to $\ell=1$, $m_{1}=0$, $m_{2}=1$, and hence this
configuration is the non-normalizable solution
of the $\Delta=4$ mode, and so describes a marginal deformation of the dual theory.

%%%%
%%%%
\section{\label{sec:chiral_mesons}Chiral Mesons from
D7-branes in AdS}
%%%%
%%%%

Just as in the flat space case, we can induce chirality into the dual
theory through the introduction of a supersymmetric magnetic
flux~\eqref{eq:mag_flux}.  However, this magnetic flux will respect
only four of the gravity supercharges, and the other four,
corresponding to the superconformal charges of the dual theory, will
not be preserved.  As we shall see, this change has important physical
consequences: the calculation of correlation functions will turn out
to require counterterms that are super-exponentially sensitive to the
ultraviolet completion of the geometry.  At the same time, the
magnetic flux induces a large amount of D3-brane charge, so that the
geometry must be sharply modified in the ultraviolet.  In practical
terms, this dependence on the ultraviolet behavior presents an
obstacle to the calculation of correlation functions.  More
importantly, it signifies that the magnetization~\eqref{eq:mag_flux},
and the corresponding appearance of chiral mesons, entails a
substantial change in the background.

\subsection{Setup and equations of motion}

We first sketch out the argument regarding the supercharges. A probe
D7-brane will preserve the supersymmetry parameterized by a Killing
double spinor $\ep$ if (cf.~\eqref{eq:D7_proj})
\begin{equation}
  P_{-}^{\uD 7}\ep=0,  \label{killing}
\end{equation}
where, with the presence of a magnetic flux $F_{2}$, $\Ga_{\uD
  7}=-\ui\Ga_{\left(8\right)}L\left(F\right)$ with
\begin{equation}
  L\bigl(F\bigr)=\sqrt{\frac{\det\left(\hat{g}\right)}
    {\det\left(\hat{g}+\ell_{\us}^{2}F\right)}}
  \biggl\{1+\frac{\ell_{\us}^{2}}{2}
  F_{\alpha_{1}\alpha_{2}}\hat{\Ga}^{\alpha_{1}\alpha_{2}}
  +\frac{\ell_{\us}^{4}}{8}
  F_{\alpha_{1}\alpha_{2}}F_{\alpha_{3}\alpha_{4}}
  \hat{\Ga}^{\alpha_{1}\alpha_{2}\alpha_{3}\alpha_{4}}\biggr\}.
\end{equation}
The bulk geometry respects the supersymmetry generated by a GKP-like
Killing spinor~\cite{Giddings:2001yu}, which is independent of the
Minkowski coordinates and annihilated by holomorphic $\ga$-matrices.
Moreover, such a Killing spinor obeys~\eqref{killing} if $F_{2}$ is $\left(1,1\right)$ and self-dual: the
$\slashed{F}_{2}$ term annihilates the Killing spinor, and the $1$ and
$\slashed{F}_{2}^{2}$ terms together are canceled by
$\sqrt{\det\left(g+F\right)}$ (which takes a simple form because $F$
is self-dual). However, the bulk geometry also supports Killing
spinors that depend on the Minkowski coordinates in a particular way (see,
e.g.,~\cite{Shuster:1999zf}).  The existence of such spinors is a
special feature of anti-de Sitter space, and the supersymmetry
transformations they induce are dual to superconformal
transformations.  Since the special AdS Killing spinors are not
preserved by the magnetized D7-brane configuration, we anticipate that
conformality will be lost in the dual theory, even in the probe
approximation.

We can also understand the loss of conformality from another point of
view.  The magnetization that gives rise to chirality follows from the
connection~\eqref{eq:mag_connection} which,
using~\eqref{eq:beta_coords}, can be written as
\begin{equation}
  A_{1}=Mr^{2}\bigl\{-\cos^{2}\beta\,\ud\phi_{1}+
  \sin^{2}\beta\,\ud\phi_{2}\bigr\},
\end{equation}
in which we are still taking $M>0$ for simplicity of presentation.
Writing $A_{1}=Mr^{2}\omega$, $\omega$ satisfies the defining equation
of a transverse vector spherical harmonic $\varrho$, which takes the
general form
\begin{equation}
  \label{eq:vec_harmonics}
  \breve{\nabla}^{2}\varrho_{\theta}=-\bigl[\ell\bigl(\ell+2\bigr)-1\bigr]
  \varrho_{\theta},\quad
  \breve{\nabla}^{\theta}\varrho_{\theta}=0,\quad
  \breve{\ep}^{\theta\vp\psi}\breve{\nabla}_{\vp}
  \varrho_{\psi}=\pm\bigl(\ell+1\bigr)
  \breve{g}^{\theta\psi}\varrho_{\psi},
\end{equation}
in which $\breve{g}$ is the metric~\eqref{eq:S3_metric} on the unit
$S^{3}$, $\breve{\nabla}$ is the associated Levi-Civita connection,
and $\breve{\ep}$ is the associated volume form.  The mode $\omega$
corresponds to the specific case $\ell=1$, with the positive sign
taken in the third equation in~\eqref{eq:vec_harmonics}.\footnote{This
  sign is independent of the sign in the equation of motion for the
  bifundamental modes,~\eqref{eq:flat_dirac}.}  Thus, $\omega$ is a
transverse vector spherical harmonic~\cite{Chodos:1983zi}, and upon
dimensional reduction leads to a canonically normalized field with
mass $m^{2}L^{2}=12$ (see~\cite{Kruczenski:2003be}).  This corresponds
to an operator of dimension $6$, and $A_{1}$ involves the
non-normalizable solution.  Hence, the introduction of the magnetic
flux deforms the dual theory by an irrelevant operator.  This implies
that not only is conformality lost in the dual theory, but the theory
does not even flow from an ultraviolet fixed point.

The addition of this flux modifies the zero-momentum equations to
\begin{subequations}
\begin{align}
  0=&
  \bigl(\bar{\partial}_{\bar{1}}\mp M z^{1}-
  \frac{1}{2}\bar{\partial}_{\bar{1}}\cA\bigr)
  \psi_{1}^{\pm}
  -\bigl(\bar{\partial}_{\bar{2}}\pm Mz^{2}
  -\frac{1}{2}\bar{\partial}_{\bar{2}}\cA\bigr)
  \psi_{2}^{\pm}
  \mp \ui q \ue^{-2\cA}z^{2}\psi_{3}^{\pm},\\
  0=&
  \bigl({\partial}_{{2}}\mp M \bar{z}^{\bar{2}}
  +\frac{3}{2}\partial_{2}\cA\bigr)\psi_{3}^{\pm}
  \mp \ui q\ue^{-2A} \bar{z}^{\bar{2}}\psi_{2}^{\pm},\\
  0=&
  \bigl({\partial}_{{1}}\pm M \bar{z}^{\bar{1}}
  +\frac{3}{2}\partial_{1}\cA\bigr)\psi_{3}^{\pm}
  \pm \ui q \ue^{-2A}\bar{z}^{\bar{2}}\psi_{1}^{\pm},\\
  0=&\bigl({\partial}_{{1}}\pm M \bar{z}^{\bar{1}}-
  \frac{1}{2}\partial_{1}\cA\bigr)\psi_{2}^{\pm}
  +\bigl({\partial}_{{2}}\mp M \bar{z}^{\bar{2}}-
  \frac{1}{2}\partial_{2}\cA\bigr)\psi_{1}^{\pm}.
\end{align}
\end{subequations}
Using~\eqref{eq:warp_factor_factor}, we find
\begin{equation}
  \label{eq:mag_warped_eom}
  0=\biggl\{\partial_{1}\bar{\partial}_{\bar{1}}+
  \partial_{2}\bar{\partial}_{\bar{2}}
  \pm M\bigl(\bar{z}^{\bar{1}}\bar{\partial}_{\bar{1}}
  -z^{1}\partial_{1}
  -\bar{z}^{2}\bar{\partial}_{\bar{2}}
  +z^{2}\partial_{2}\bigr)
  -M^{2}\bigl\lvert z^{1}\bigr\rvert^{2}
  -\bigl(M^{2}+\ue^{-4\cA}q^{2}\bigr)\bigl\lvert z^{2}\bigr\rvert^{2}
  \biggr\}\vp_{\pm},
\end{equation}
where $\vp_{\pm}=\frac{1}{\bar{z}^{\bar{2}}}\vp_{\pm}^{3}$.  With the
coordinates~\eqref{eq:beta_coords}, this becomes
\begin{multline}
  \label{eq:mag_psi_2}
  0=\biggl\{
  \partial_{r}^{2}
  +\frac{3}{r}\partial_{r}
  \pm 4\ui M\bigl(\partial_{\phi_{1}}-\partial_{\phi_{2}}\bigr)
  -4M^{2}r^{2}\\
  +\frac{1}{r^{2}}\biggl[
  \partial_{\beta}^{2}
  +\bigl(\cot\beta-\tan\beta\bigr)\partial_{\beta}
  +\frac{1}{\cos^{2}\beta}\partial_{\phi_{1}}^{2}
  +\frac{1}{\sin^{2}\beta}\partial_{\phi_{2}}^{2}
  \biggr]
  -\frac{4q^{2}r^{2}L^{4}\sin^{2}\beta}{\left(r^{2}+\mu^{2}\right)^{2}}
  \biggr\}\vp_{\pm}.
\end{multline}
Again, the relative simplicity of this equation is a consequence of
the self-duality constraint imposed by supersymmetry.  When $\mu=0$,
the equation is again separable and it is useful to take the
ansatz~\eqref{eq:split_ansatz}.  $Q_{\pm}$ satisfies the same
eigenvalue problem~\eqref{eq:angular} while the radial equation is now
\begin{equation}
  0=f_{\pm}''+\frac{3}{r}f_{\pm}'\mp 4M\bigl(m_{1}-m_{2}\bigr)f_{\pm}
  -4M^{2}r^{2}f_{\pm}-\frac{\lambda}{r^{2}}f_{\pm}.
\end{equation}
The solutions can be expressed in terms of $\cM$ and $\cU$, the confluent
hypergeometric functions of the first and second kind,
\begin{equation}
  \label{eq:magnetized_soln}
  f_{\pm}=\ue^{-Mr^{2}}r^{-\nu}
  \biggl\{c_{1}\,
  \cM\bigl(\mu;\nu;2M r^{2}\bigr)
  + c_{2}\,\cU\bigl(\mu;\nu;2M r^{2}\bigr)\biggr\},
\end{equation}
in which
\begin{equation}
  \nu=1+\sqrt{1+\lambda} \qquad {\rm{and}} \qquad
  \mu=\frac{1}{2}\bigl(\nu\pm\bigl(m_{1}-m_{2}\bigr)\bigr).
\end{equation}

As anticipated, the solutions are not power
laws, and so the dual field theory is no longer conformal even in the
probe approximation.  Furthermore, noting that the  dominant asymptotic behavior
at $r \to \infty$ is
\begin{equation}
\cM\bigl(\mu;\nu;2M r^{2}\bigr) \propto e^{2M r^2},
\end{equation} where we have omitted power law factors,
we find that the divergent part
of~\eqref{eq:magnetized_soln} grows super-exponentially at $r \to \infty$:
\begin{equation}
  \label{eq:magnetized_soln_asym}
  f_{\pm} \propto e^{M r^2} .
\end{equation}

\subsection{Ultraviolet sensitivity  of the correlation functions}

To interpret the divergences identified above, it will be helpful to
recall the well-established procedure for computing correlation
functions in AdS/CFT, focusing on the process of removing divergences
of the classical action through the introduction of counterterms,
i.e.~holographic renormalization (see~\cite{Skenderis:2002wp} for a
review).

The basic statement of the duality, in the limit~\eqref{decouplinglimit}, is the
identification of the generating functional of the CFT with the
classical supergravity action,
\begin{equation}
  \mathcal{Z}_{\mathrm{CFT}}=\ue^{-S_{\mathrm{grav}}}.
\end{equation}
An operator $\cO$ on the field theory side has a corresponding
classical field $\vp$ on the gravity side.  If $\cO$ is a scalar
field, then $\vp$ also transforms as an $\SO{3,1}$ scalar.  The
solution for $\vp$ at large $r$ can be separated into a dominant term
and a subdominant term,
\begin{equation}
  \vp=a_{\mathrm{dom}}\vp_{\mathrm{dom}}+a_{\mathrm{sub}}\vp_{\mathrm{sub}}.
\end{equation}
If the geometry is asymptotically anti-de Sitter space, both the
dominant and subdominant terms are power laws at large $r$.  Moreover,
$a_{\mathrm{dom}}$ is dual to a source term for $\cO$, and correlation
functions of $\cO$ are calculated by
taking functional
derivatives of $S_{\mathrm{grav}}$ with respect to $a_{\mathrm{dom}}$
and then later taking $a_{\mathrm{dom}}\to 0$.

For finite $a_{\mathrm{dom}}$, the classical action
$S_{\mathrm{grav}}$ is divergent.  This can be addressed by adding
counterterms to the action: one first regulates the action by cutting
off the space at a large but finite radius $r_{\Lambda}$.  The terms
that diverge as $r_{\Lambda}\to\infty$ are canceled by adding terms to
the supergravity action that are localized on the boundary at
$r_{\Lambda}$.  Taking $r_{\Lambda}\to\infty$ then yields a finite
action.  The power law behavior of solutions in the AdS case means
that such counterterms have power-law (and potentially logarithmic)
dependence on $r_{\Lambda}$.  However, the super-exponential
growth~\eqref{eq:magnetized_soln_asym} of the chiral modes requires
the introduction of counterterms that have a similar super-exponential
dependence on the cutoff.  Since the magnetization required to induce
chirality deforms the theory by an irrelevant operator, such strong
sensitivity is perhaps not surprising.

If
the background
remained
unaltered by magnetization, the structure of counterterms would
represent a technically demanding but potentially surmountable
challenge to calculating correlation functions.\footnote{For example,
  as developed in~\cite{Krasnitz:2002ct,*Aharony:2005zr}, it is
  possible to calculate correlation functions in the KT/KS
  theory~\cite{Klebanov:2000nc,Klebanov:2000hb}, even though the
  theory does not flow from an ultraviolet fixed point.} However, the
chirality-inducing magnetic flux sources a large amount of D3-brane
charge via the Chern-Simons coupling $\int C_{4}\wedge F_{2}\wedge
F_{2}$:
the dissolved D3-brane flux diverges as
\begin{equation}
  \label{eq:induced_charge}
  \int_{R<\rho}F_{2}\wedge F_{2}\sim M^{2} \rho^{4}.
\end{equation}
This is comparable to the D3-brane charge of the background when
\begin{equation}
  \rho\sim\frac{N^{1/4}}{M^{1/2}},
\end{equation}
at which point the influence of this charge on the geometry must be
taken into account.  A calculation of correlation functions that fails
to incorporate this backreaction is not physically meaningful.

One might ask whether  a different choice of magnetization (still without a localized source) results in a different conclusion.
Supersymmetric fluxes supported on the D7-branes are characterized by scalar hyperspherical harmonics
--- cf.~\eqref{bianchi} --- and so the
fluxes grow as $F_{2}\sim
r^{j+2}\Omega^{\left(j\right)}+r^{j+1}\ud
r\wedge\omega^{\left(j\right)}$, where $j=0,1,2,\ldots$, and
$\Omega^{\left(j\right)}$ and $\omega^{\left(j\right)}$ are a $2$-form
and a $1$-form on $S^{3}$, respectively.
Our
analysis of the magnetic flux~\eqref{magnetizationchoice} corresponds
to the case $j=0$.  Other values of $j$ would lead to steeper
potentials in~\eqref{eq:mag_warped_eom}, and so to a greater degree of
localization of the bifundamental wavefunctions.
However, the charge carried by such flux diverges more quickly than~\eqref{eq:induced_charge},
growing as $r^{2j+4}$, and hence the problem of ultraviolet sensitivity is exacerbated.

%%%%
%%%%
\section{\label{sec:conc}Conclusions}
%%%%
%%%%

In this note we analyzed the spectrum of mesonic operators arising
from strings stretching between intersecting D7-branes in
$AdS_{5}\times S^{5}$.  The dual field theory is an $\cN=1$ deformation of
maximally supersymmetric $\SU{N}$ SYM, with the addition
of a $\U{F_{1}}\times\U{F_{2}}$ flavor group, under which the $7$-$7'$
strings transform as bifundamentals.\footnote{For notational
  simplicity only, we limited our discussion to the case
  $F_{1}=F_{2}=1$, corresponding to a single pair of D7-branes.}  We
considered D7-branes with and without magnetic flux on the curve of
intersection, finding sharply different results in these two cases.

The intersection of the D7-branes corresponds to a particular adjoint
Higgsing of the $\U{2}$ theory arising on coincident D7-branes.  In
the field theory, the fact that the branes intersect is described by a
marginal deformation.  If the D7-branes reach the origin of warping,
and one furthermore makes the quenched/probe approximation that
neglects backreaction of the D7-branes, then the dual theory is
conformal.  In this case --- where magnetization has not yet been
incorporated --- we computed the spectrum of dual operators.  The
$7$-$7'$ strings are mixtures of the transverse deformations and the
internal components of the gauge field, and as a consequence the
equations of motion are difficult to solve analytically.  However,
conformal symmetry leads to a remarkable simplification of the
equations of motion, through which we were able to find numerical
solutions.  The behavior of the dimensions depends on the value of
$\xi \sim \tan\theta\sqrt{ g_{\us}N}$ , cf.~\eqref{xidef}, where
$\theta$ is an angle characterizing the intersection.  Approximate
spectra are given in~\eqref{eq:spectrum}.  As expected, the modes are
well localized along the intersection of the D7-branes and have
power-law behavior along the holographic direction.

We then considered introducing magnetic flux on the curve of
intersection, leading to a chiral spectrum in the dual theory.  The
simplest magnetization corresponds to an irrelevant deformation of the
theory, by an operator of dimension $\Delta=6$.  As a consequence, the
non-normalizable solutions to the bifundamental equations of motion
have super-exponential divergence in the ultraviolet,
cf.~\eqref{eq:magnetized_soln_asym}.  Although the
limit~\eqref{decouplinglimit} allows us to neglect the backreaction of
the D7-branes themselves, the backreaction of the D3-brane charge
induced by the magnetic flux cannot be neglected.  Since the
calculation of correlation functions, for example through holographic
renormalization, requires the use of the non-normalizable modes, the
procedure for calculating the correlation functions is unclear.  This
is a physical limitation rather than a technical one: the divergence
of the D3-brane charge induced by magnetization of noncompact
D7-branes signals the need for an ultraviolet completion via
compactification.  In the dual language, the field theory describing
magnetized D7-branes does not flow from an ultraviolet fixed point.

On the other hand, we found that the normalizable modes of the chiral
bifundamental mesons are very well localized in the infrared.
Indeed, at large $r$,
\begin{equation}
  \cU\bigl(\mu;\nu;2Mr^{2}\bigr)\sim r^{-\mu},
\end{equation}
so that, when $c_{1}=0$ in~\eqref{eq:magnetized_soln}, the bifundamental
modes exhibit a Gaussian localization,
\begin{equation}
  f_{\pm}\propto \ue^{-Mr^{2}} ,
\end{equation}
where we have again omitted power law factors and have chosen $M>0$.
Although similar Gaussian peaks appear in flat space (see
e.g.~\cite{Cremades:2004wa}), this feature in warped space has the
potential to provide a rich playground for model-building.  In
general, the lack of knowledge of the metric and of related fields
often stymies detailed model-building in string compactifications.
However, the metrics for infinite families of non-compact (and
singular) Calabi-Yau cones are known explicitly.  These cones can be
used to construct strongly warped geometries that can be attached to
compact spaces --- see for example the discussion
in~\cite{Giddings:2001yu}.  Attachment to a compactification modifies
the solution in the cone region, by introducing sources for irrelevant
perturbations, but these effects can be incorporated systematically,
as in~\cite{Baumann:2010sx,*Gandhi:2011id}.  One can therefore build a
local model on D3-branes at the apex of the cone, but also take into
account bulk effects, including supersymmetry breaking and moduli
stabilization.  Constructions in this corner of the landscape are
limited to some degree by the possible singularities at the apex.
An alternative, toward which the present work is a modest
advance, is to consider model-building on intersecting magnetized
D7-branes.  Although the D7-branes will stretch beyond the warped
region into the bulk,\footnote{Indeed, the consistency of embeddings
  in global models will provide constraints on which models can be
  built.} we have demonstrated that at least some bifundamental modes
are well localized in the infrared.  This allows for a combination of
the richness of model-building with intersecting D7-branes and the
power of local model-building in warped geometries.  Although we
limited our particular analysis to $AdS_{5}\times S^{5}$, the
qualitative result should extend to more general cones and their
deformations (though the details, of course, become much more
complex).

This localization also implies that although correlation functions are
difficult to describe, the mass spectrum of mesons can in principle be
calculated with reliable numerical techniques.  When the
D7-branes move away from the center of $AdS_{5}$, the spectrum of
mesons becomes gapped even though, in the quenched approximation, the
glueball spectrum is continuous~\cite{Kruczenski:2003be}.  A standard
method of finding the meson mass spectrum in the gapped case is to calculate the
correlation functions and check for the appearance of poles.  However,
a practical alternative is to find those solutions that satisfy
appropriate infrared boundary conditions and are normalizable in the
ultraviolet
(see, for example,~\cite{Kruczenski:2003be,Berg:2006xy}).
Because the equation of motion constitutes a Sturm-Liouville problem,
this alternative approach leads to a discrete spectrum, and since the
solutions are expected to be exponentially convergent, the resulting
spectrum would be reliable.  On the other hand, once the spectrum becomes
gapped the radial and angular parts of the equation of motion no
longer separate, even in the unmagnetized case~\eqref{eq:warped_psi_2}.
This is a significant complication, and so we leave this analysis to
future work.

Yet another possibility is to consider alternative magnetizations.
The magnetization that we analyzed in this note is the simplest
unsourced magnetic flux that is possible in our construction, and
other unsourced magnetic fluxes would enhance the bifundamental
wavefunction localization that we found, while intensifying the
problem of ultraviolet sensitivity.  Magnetic flux that is itself
localized in the infrared, and produces only normalizable
perturbations to the geometry, would require a local source.  In
particular, it was pointed out in~\cite{Benini:2009ff} and explicitly
shown in~\cite{McGuirk:2012tv} that the addition of anti-D3-branes to
warped flux backgrounds provides an infrared-localized magnetization.
Although the resulting magnetization has a gauge structure that
differs from (\ref{magnetizationchoice}) --- specifically, the induced
magnetization is proportional to the identity --- this remains an
intriguing possibility for future work.

%%%%
%%%%
\acknowledgments
%%%%
%%%%

It is a pleasure to thank F.~Marchesano and G.~Shiu for useful
discussions of related topics.  This work was supported by the NSF
under grant PHY-0757868.

\appendix

%%%%
%%%%
\section{\label{app:conv}Conventions for Fermions}
%%%%
%%%%

In this appendix we summarize our conventions for fermions, many of
which follow from~\cite{Polchinski:1998rr}.  We work with a Weyl basis
for the $\SO{9,1}$ $\Ga$-matrices and make use of the decomposition
$\SO{9,1}\to\SO{3,1}\times\SO{6}$.  For $\SO{3,1}$ we take
\begin{equation}
  \ga^{0}=\begin{pmatrix}
     & \II_{2} \\ -\II_{2} &
  \end{pmatrix},\quad
  \ga^{i=1,2,3}=\begin{pmatrix}
     & \sig^{i} \\ \sig^{i} &
  \end{pmatrix},
\end{equation}
in which $\sig^{i}$ are the Pauli matrices
\begin{equation}
  \sig^{1}=\begin{pmatrix} 0 & 1\\ 1 & 0\end{pmatrix},\quad
  \sig^{2}=\begin{pmatrix} 0 & -\ui \\ \ui & 0\end{pmatrix},\quad
  \sig^{3}=\begin{pmatrix} 1 & 0 \\ 0 & -1\end{pmatrix}.
\end{equation}
For $\SO{2k+1,1}$, we take the chirality matrix to be
\begin{equation}
  \ga_{\left(2k+2\right)}=\ui^{-k}\ud\slashed{\vol}_{\mathbb{R}^{2k+1,1}},
\end{equation}
where $\ud\vol_{M}$ is the volume element on $M$
\begin{equation}
  \ud\vol_{M}=\frac{1}{d!}\ep_{M_{1}\cdots M_{d}}\ud x^{M_{1}}
  \wedge\cdots\wedge\ud x^{M_{d}},
\end{equation}
in which $\ep_{01\cdots\left(d-1\right) }=\sqrt{-\det g}$.  For $\mathbb{R}^{3,1}$,
\begin{equation}
  \ga_{\left(4\right)}=
  -\ui\ga^{0}\ga^{1}\ga^{2}\ga^{3}
  =\begin{pmatrix}\II_{2} & \\ & -\II_{2}\end{pmatrix}.
\end{equation}
The 4d Majorana matrix is
\begin{equation}
  \beta_{4}=\ga_{\left(4\right)}\ga^{2}
  =\begin{pmatrix} & -\sig^{2} \\ \sig^{2} & \end{pmatrix}.
\end{equation}

For $\SO{6}$, we define
\begin{align}
  \tilde{\ga}^{4}=&\sig^{1}\otimes\II_{2}\otimes\II_{2},&
  \tilde{\ga}^{7}=&\sig^{2}\otimes\II_{2}\otimes\II_{2},\notag\\
  \tilde{\ga}^{5}=&\sig^{3}\otimes\sig^{1}\otimes\II_{2},&
  \tilde{\ga}^{8}=&\sig^{3}\otimes\sig^{2}\otimes\II_{2},\notag\\
  \tilde{\ga}^{6}=&\sig^{3}\otimes\sig^{3}\otimes\sig^{1},&
  \tilde{\ga}^{9}=&\sig^{3}\otimes\sig^{3}\otimes\sig^{2}.\notag
\end{align}
For $\SO{2k+4}$, the chirality operator is
\begin{equation}
  \ga_{\left(2k+4\right)}=\ui^{-k}\ud\slashed{\vol}_{\mathbb{R}^{2k+4}},
\end{equation}
and so
\begin{equation}
  \tilde{\ga}_{\left(6\right)}
  =-\ui\,\tilde{\ga}^{1}\cdots\tilde{\ga}^{6}
  =\sig^{3}\otimes\sig^{3}\otimes\sig^{3}.
\end{equation}
The Majorana matrix is
\begin{equation}
  \tilde{\beta}_{6}=\tilde{\ga}^{7}\tilde{\ga}^{8}\tilde{\ga}^{9}
  =\sig^{2}\otimes\ui\,\sig^{1}\otimes\sig^{2}.
\end{equation}
We will make use of a complex structure
\begin{equation}
  z^{I}=x^{3+I}+\ui\,x^{4+I}.
\end{equation}
Defining
\begin{equation}
  \sig^{\pm}=\frac{1}{2}\bigl(\sig^{1}\pm\ui\,\sig^{2}\bigr),
\end{equation}
we have
\begin{align}
  \tilde{\ga}^{1}=&2\,\sig^{+}\otimes\II_{2}\otimes\II_{2},&
  \tilde{\ga}^{\bar{1}}=&2\,\sig^{-}\otimes\II_{2}\otimes\II_{2},\notag\\
  \tilde{\ga}^{2}=&2\,\sig^{3}\otimes\sig^{+}\otimes\II_{2},&
  \tilde{\ga}^{\bar{2}}=&2\,\sig^{3}\otimes\sig^{-}\otimes\II_{2},\notag\\
  \tilde{\ga}^{3}=&2\,\sig^{3}\otimes\sig^{3}\otimes\sig^{+},&
  \tilde{\ga}^{\bar{3}}=&2\,\sig^{3}\otimes\sig^{3}\otimes\sig^{-}.\notag
\end{align}
We can construct a basis of positive chirality spinors by first
defining
\begin{equation}
  \eta_{+}=\begin{pmatrix} 1 \\ 0\end{pmatrix},\quad
  \eta_{-}=\begin{pmatrix} 0 \\ 1\end{pmatrix}.\quad
\end{equation}
The positive chirality spinors are then
\begin{equation}
  \label{eq:pos_basis}
  \eta_{0}=\eta_{+++},\quad\eta_{1}=\eta_{+--},
  \quad\eta_{2}=\eta_{-+-},\quad\eta_{3}=\eta_{--+},
\end{equation}
in which
\begin{equation}
  \eta_{\ep_{1}\ep_{2}\ep_{3}}=\eta_{\ep_{1}}\otimes
  \eta_{\ep_{2}}\otimes
  \eta_{\ep_{3}}.
\end{equation}
Note that $\sig^{\pm}\eta_{\pm}=0$, so that $\eta_{+++}$ is annihilated
by all contravariant holomorphic $\tilde{\ga}$-matrices.

Finally, we construct the $\SO{9,1}$ $\Ga$-matrices by
\begin{equation}
  \hat{\Ga}^{\mu}=\ga^{\mu}\otimes\II_{8},\quad
  \hat{\Ga}^{m}=\ga_{\left(6\right)}\otimes\tilde{\ga}^{m}.
\end{equation}
The chirality and Majorana matrices are
\begin{align}
  \hat{\Ga}_{\left(10\right)}=&\hat{\Ga}^{0}\hat{\Ga}^{1}\cdots\hat{\Ga}^{9}
  =-{\ga}_{\left(4\right)}\otimes\tilde{\ga}_{\left(6\right)},\notag\\
  \hat{B}_{10}=&\hat{\Ga}^{2}\hat{\Ga}^{7}\hat{\Ga}^{8}\hat{\Ga}^{9}
  =-\beta_{4}\otimes\tilde{\beta}_{6}.
\end{align}

We will make use of 32-component Majorana-Weyl spinors satisfying
\begin{equation}
  \label{eq:MW_cond}
  \hat{\Ga}_{\left(10\right)}\theta=-\theta,\quad
  \hat{B}_{10}\theta=\theta^{\ast}.
\end{equation}
An example of such a spinor is
\begin{equation}
  \theta=
  \begin{pmatrix}
    \xi \\ 0
  \end{pmatrix}\otimes\eta
  -\begin{pmatrix}
    0 \\ \sig^{2}\xi^{\ast}
  \end{pmatrix}
  \otimes\tilde{\beta}_{6}\eta^{\ast},
\end{equation}
where $\tilde{\ga}_{\left(6\right)}\eta=+\eta$.

We will also make use of double spinors built from pairs of 10d
Majorana-Weyl spinors
\begin{equation}
  \Theta=\begin{pmatrix}\theta^{1} \\ \theta^{2}\end{pmatrix},
\end{equation}
where both $\theta^{1}$ and $\theta^{2}$ satisfy~\eqref{eq:MW_cond}.
$\hat{\Ga}$-matrices act on double spinors as
\begin{equation}
  \hat{\Ga}^{M}\Theta=\begin{pmatrix} \hat{\Ga}^{M}\theta^{1} \\
    \hat{\Ga}^{M}\theta^{2}\end{pmatrix},
\end{equation}
while explicit Pauli matrices act to mix the elements of the double
spinor.  For example,
\begin{equation}
  \sig^{1}\begin{pmatrix}\theta^{1} \\ \theta^{2}\end{pmatrix}
  =\begin{pmatrix}\theta^{2} \\ \theta^{1}\end{pmatrix}.
\end{equation}

%%%%
%%%%
\section{\label{app:spherical}Hyperspherical Harmonics}
%%%%
%%%%

In this appendix we review a few properties of the hyperspherical
harmonics on $S^{3}$.  A useful parametrization of $S^{3}$ is via the
usual embedding of $S^{3}$ into $\mathbb{R}^{4}$,
$\zeta^{i}\zeta^{i}=1$, where $\zeta^{1}\ldots \zeta^{4}$ are
coordinates on $\mathbb{R}^{4}$.  We take (as in, for
example,~\cite{Meremianin})
\begin{align}
  \zeta^{1}=&r\cos\beta\cos\phi_{1},
  &\zeta^{2}=&r\sin\beta\cos\phi_{2},\notag\\
  \zeta^{3}=&r\cos\beta\sin\phi_{1},
  &\zeta^{4}=&r\sin\beta\sin\phi_{2},
\end{align}
with $\beta\in\left[0,\frac{\pi}{2}\right]$ and
$\phi_{a}\in\left[0,2\pi\right)$. The induced metric on $S^{3}$ is
\begin{equation}
  \label{eq:S3_metric}
  \ud s_{S^{3}}^{2}=\breve{g}_{\theta\vp}\ud y^{\theta}\ud y^{\vp}
  =\ud\beta^{2}+\cos^{2}\beta\,\ud\phi_{1}^{2}
  +\sin^{2}\beta\,\ud\phi_{2}^{2}.
\end{equation}
The volume of $S^{3}$ is
\begin{equation}
  \cV_{S^{3}}=\int_{0}^{2\pi}\ud\phi_{1}\int_{0}^{2\pi}\ud\phi_{2}
  \int_{0}^{{\pi}/2}\ud\beta\sin\beta\cos\beta=2\pi^{2}.
\end{equation}
The scalar spherical harmonics satisfy the eigenvalue problem
\begin{equation}
  \breve{\nabla}^{2}\cY=\frac{\partial^{2}\cY}{\partial\beta^{2}}
  +\bigl(\cot\beta-\tan\beta\bigr)\frac{\partial\cY}{\partial\beta}
  +\frac{1}{\cos^{2}\beta}\frac{\partial^{2}\cY}{\partial\phi_{1}^{2}}
  +\frac{1}{\sin^{2}\beta}\frac{\partial^{2}\cY}{\partial\phi_{2}^{2}}
  =-\lambda\cY.
\end{equation}
Taking the ansatz
\begin{equation}
  \cY=\ue^{\ui\left(m_{1}\phi_{1}+m_{2}\phi_{2}\right)}y\bigl(\cos2\beta\bigr)
\end{equation}
gives
\begin{equation}
  0=4\bigl(1-x^{2}\bigr)y''-8xy'-\frac{2m_{1}^{2}}{1+x}y
  -\frac{2m_{2}^{2}}{1-x}y
  +\lambda y,
\end{equation}
in which $x=\cos 2\beta$.  Imposing Neumann conditions so that a zero
mode is admitted, the solutions are given in terms of Jacobi polynomials
$P_{r}^{\left(a,b\right)}$,
\begin{equation}
  \cY_{\ell,m_{1},m_{2}}\bigl(\beta,\phi_{1},\phi_{2}\bigr)
  =c_{\ell,m_{1},m_{2}}\ue^{\ui\left(m_{1}\phi_{1}+m_{2}\phi_{2}\right)}
  \bigl(1+\cos 2\beta\bigr)^{m_1/2}
  \bigl(1-\cos 2\beta\bigr)^{m_2/2}
  P_{r}^{\left(m_{2},m_{1}\right)}
  \bigl(\cos 2\beta\bigr),
\end{equation}
in which $r=\frac{1}{2}\left(\ell-m_{1}-m_{2}\right)$.  For these to
be non-vanishing regular solutions, $r$ must be an integer and
\begin{equation}
  \label{eq:spherical_harm_requirement}
  0\le\bigl\lvert m_{1}\bigr\rvert+\bigl\lvert m_{2}\bigr\rvert
  \le\ell.
\end{equation}
These solutions satisfy
\begin{equation}
  \breve{\nabla}^{2}\cY=-\ell\left(\ell+2\right)\cY,
\end{equation}
and the condition~\eqref{eq:spherical_harm_requirement} gives the
expected degeneracy of $\left(\ell+1\right)^{2}$ (see, for
example,~\cite{Chodos:1983zi}).

The Jacobi polynomials are orthogonal in the sense that
\begin{equation}
  \int_{-1}^{1}\ud x\,
  \bigl(1-x\bigr)^{a}\bigl(1+x\bigr)^{b}
  P_{r}^{\left(a,b\right)}P_{s}^{\left(a,b\right)}
  =\frac{2^{a+b+1}}{2r+a+b+1}
  \frac{\left(a+r\right)!\,\left(b+r\right)!}
  {r!\,\left(a+b+r\right)!}\delta_{rs}.
\end{equation}
Therefore the normalization condition
\begin{equation}
  \int\ud\vol_{S^{3}}\,\cY^{\ast}_{\ell,m_{1}^{\phantom{,}},m_{2}^{\phantom{,}}}
  \cY^{\phantom{\ast}}_{\ell',m'_{1},m'_{2}}=
  \delta_{\ell'\ell_{\phantom{,\!}}\phantom{m_{1}'m_{1}^{\phantom{,\!}}}}
  \hspace{-2.1em}
  \delta_{m_{1}'m_{1}^{\phantom{,\!}}\phantom{\ell'\ell_{\phantom{,\!}}}}
  \hspace{-1.1em}
  \delta_{m_{2}'m_{2}^{\phantom{,\!}}\phantom{\ell'\ell_{\phantom{,\!}}}}
  \hspace{-1.1em}
\end{equation}
is satisfied by taking
\begin{equation}
  \label{eq:norm}
  c_{\ell,m_{1},m_{2}}=\frac{1}{\pi}
  \sqrt{\frac{\ell+1}{2^{m_{1}+m_{2}+1}}
    \frac{\left[\frac{1}{2}\left(\ell+m_{1}+m_{2}\right)\right]!
      \left[\frac{1}{2}\left(\ell-m_{1}-m_{2}\right)\right]!}
    {\left[\frac{1}{2}\left(\ell+m_{1}-m_{2}\right)\right]!
      \left[\frac{1}{2}\left(\ell-m_{1}+m_{2}\right)\right]!}}.
\end{equation}

The Jacobi polynomials satisfy the useful recursion relationship
\begin{multline}
  \label{eq:recursion}
  x\,P^{\left(a,b\right)}_{r}\left(x\right) =
  \frac{2\left(a+r\right)\left(b+r\right)}
  {\left(a+b+2r\right)\left(a+b+2r+1\right)}
  P^{\left(a,b\right)}_{r-1}\left(x\right)\\
  +\frac{2\left(r+1\right)\left(a+b+r+1\right)}
  {\left(a+b+2r+1\right)\left(a+b+2r+2\right)}
  P^{\left(a,b\right)}_{r+1}\left(x\right)
  +\frac{b^{2}-a^{2}}
  {\left(a+b+2r\right)\left(a+b+2r+2\right)}
  P^{\left(a,b\right)}_{r}\left(x\right).
\end{multline}

\bibliography{mms}

\end{document}